\pdfoutput=1
\documentclass[12pt]{article}

\usepackage[margin=1in]{geometry}
\usepackage[x11names,dvipsnames]{xcolor}
\usepackage{amsmath,amssymb}
\usepackage{graphicx}
\usepackage{float}
\usepackage{subfig}
\usepackage{scrextend}
\usepackage[T1]{fontenc}
\usepackage{lmodern} 

\usepackage{booktabs}
\usepackage{longtable}
\usepackage{array}
\usepackage{multirow}
\usepackage{wrapfig}
\usepackage{makecell}
\usepackage{threeparttable}

\usepackage[numbers, square]{natbib}
\usepackage[affil-it]{authblk}

\makeatletter
\let\saved@footnotemark\@footnotemark 
\usepackage{bookmark}
\usepackage{hyperref}
\IfFileExists{xurl.sty}{\usepackage{xurl}}{} 
\hypersetup{
  colorlinks=true,
  linkcolor={blue},
  filecolor={Maroon},
  citecolor={Blue},
  urlcolor={Blue}
}
\let\@footnotemark\saved@footnotemark 
\makeatother

\title{Unified implementation and comparison of Bayesian shrinkage methods for treatment effect estimation in subgroups}

\author[1*]{Marcel Wolbers}
\author[2*]{Miriam Pedrera Gómez}
\author[3]{Alex Ocampo}
\author[1]{Isaac Gravestock}

\affil[1]{Clinical Statistics and Analytics, Bayer BCC AG, Basel, Switzerland} 
\affil[2]{Indivi AG, Basel, Switzerland} 
\affil[3]{Data Science and Analytics, Roche, Basel, Switzerland} 
\affil[*]{These authors contributed equally to this work.}

\begin{document}
\maketitle

\begin{abstract}
Evaluating treatment effect heterogeneity across patient subgroups is a fundamental aspect of clinical trial analysis. Yet, these analyses have inherent limitations due to small sample sizes and the substantial number of subgroups investigated. There is also a tendency to focus on extreme estimates, which may reflect random variation rather than true effects, potentially leading to spurious clinical conclusions. In recent years, statisticians in regulatory agencies and pharmaceutical companies have begun considering shrinkage methods grounded in Bayesian statistical theory. These methods incorporate priors on treatment effect heterogeneity, which operationally shrink raw subgroup treatment effect estimates towards the overall treatment effect. Various shrinkage estimators and priors have been proposed, yet it remains unclear which methods perform best. This work provides a unified presentation and software implementation for continuous, binary, count, and time-to-event endpoints in the \texttt{R} package \texttt{bonsaiforest2}.
It also provides simulation comparisons of one-way and global shrinkage methods for two distinct simulation set-ups. One-way models fit a separate shrinkage model for each subgrouping variable, whereas global models fit a model including all subgroup indicators at once. Both can derive standardized subgroup-specific treatment effects. Across all simulation scenarios, shrinkage methods outperformed the standard subgroup estimator without shrinkage in terms of mean squared error. They were also more efficient in identifying a non-efficacious subgroup. Global shrinkage models tended to have smaller mean squared error and less dependence on hyperprior parameters than one-way models, but also exhibited slightly larger bias and worse frequentist coverage of associated credible intervals. For both models, hyperprior choices anchored in trial assumptions about the anticipated size of the overall treatment effect performed well. We conclude that some degree of shrinkage is preferable to none and advocate for the routine inclusion of shrunken estimates in clinical forest plots to facilitate more robust decision-making.
\end{abstract}

\section{Introduction}

Subgroup analyses are an integral component of the statistical analysis of randomized controlled clinical trials.
We focus on exploratory subgroup analysis, specifically consistency assessments across a relatively small number (e.g., 10 to 30) of pre-specified subgroups \citep{lipkovich2017tutorial}.
Standard subgroup analyses typically involve visualizing treatment effect estimates and associated confidence intervals in forest plots, alongside treatment-by-subgroup interaction tests with their corresponding p-values.

Subgroup analyses are inherently challenging due to the low precision of subgroup-specific estimates, the limited power of interaction tests, the risk of multiple testing across many categories, and the tendency to over-interpret extreme estimates.
Consequently, these analyses are often treated with caution \citep{sleight2000debate}.
While strict standards have been proposed to assess the credibility of claims regarding effect modification \citep{schandelmaier2020iceman, emaSubgroups2019}, it remains difficult to determine whether the overall treatment effect applies to the full population or if specific subgroups have a substantially increased or decreased treatment effect.
An additional complication is that the assessment of treatment effect homogeneity depends on the effect scale. A further technical issue is that measures such as odds ratios and hazard ratios are non-collapsible and not ``logic-respecting,'' which may conflate prognostic and predictive variables \citep{daniel2021,liu2022}. A consequence of this is that subgroup analyses that condition on different sets of auxiliary baseline covariates target different estimands altogether for non-collapsible effect measures.

To improve the precision of treatment effect estimation, Bayesian shrinkage methods that borrow information across subgroups have been proposed in the statistical literature \citep{jones2011, henderson2016, wang2024FDA,wolbers2025shrinkage} and in regulatory guidance \citep{iche17,fdaBayesian2026}.
Empirical evidence from identical Phase III trials suggests that shrinkage methods can outperform standard estimation techniques in predictive accuracy \citep{bornkamp2025}.
Complementary simulation findings further demonstrate that these methods achieve a substantially lower mean squared error across subgroup analyses \citep{wolbers2025shrinkage}.

There are two primary approaches to Bayesian shrinkage: \emph{One-way shrinkage models} fit a Bayesian hierarchical model for one subgrouping variable at a time \citep{jones2011, henderson2016, wang2024FDA}, typically requiring multiple model fits to obtain effects for all subgroups of interest.
In contrast, \emph{global shrinkage models} fit a single model including treatment interactions for all subgroups jointly and obtain effects via standardization \citep{jones2011,dixon1991,wolbers2025shrinkage}.

A unified software implementation to fit both one-way and global shrinkage models as well as simulation studies comparing these methods are currently lacking.
This paper aims to fill this gap.
We describe the methods implemented in our R package in Section~\ref{sec:methods}, apply them to a case study in Section~\ref{sec:gallium}, and compare them via two sets of simulations in Section~\ref{sec:simulation}.
We conclude with a discussion in Section~\ref{sec:discussion}.

\section{Methods}\label{sec:methods}

We discuss methods applicable to data from an RCT of an intervention versus a control treatment.
Estimators for treatment effects in the subgroups are derived in two steps: First, we fit a Bayesian outcome model to the data.
Second, we obtain treatment effect estimates for subgroups defined by a single subgrouping variable at a time via standardization, i.e., by marginalizing over the covariate distribution within the subgroup. The methods are implemented in a flexible R package \texttt{bonsaiforest2} \citep{bonsaiforest2}.

We denote the subjects included in the trial by the subscript $i$ ($i = 1,\ldots , N)$ and their treatment assignment by $z_i$ taking values of 0 for control and 1 for intervention.
For the \emph{one-way shrinkage model}, assume that we are interested in treatment effects in subgroups defined by the levels of a single categorical subgrouping variable $x$ with $K$ levels.
For \emph{global shrinkage models}, assume that subgroups are deﬁned by the levels of $p$ categorical subgrouping variables $x_j$ $(j = 1,\ldots , p)$, respectively. Each variable $x_j$ has $l_j$ levels, resulting in a total of $K = \sum_{j=1}^p l_j$ subgroups defined by a single variable at a time. For both models, denote indicators for each of the $K$ subgroups by $s_{ik}$ (i.e., $s_{ik}=1$ if subject $i$ belongs to the subgroup $k$, and $s_{ik}=0$ otherwise, with $k=1,\ldots, K$).
In one-way models, only a single indicator $s_{ik}$ per subject is equal to 1; in global shrinkage models, $p$ indicators per subject are equal to 1 because the subgroups defined by different subgrouping variables are overlapping. In addition, the model might be adjusted for additional baseline covariates denoted by $u_{il}$ ($l=1,\ldots,L$).

\subsection{Outcome model}

The outcome model for both one-way and global shrinkage is a Bayesian regression model with a linear predictor structure.
Our implementation supports continuous, binary, count, and time-to-event endpoints, which are modeled via linear, logistic, negative binomial, and Cox regression, respectively.
The resulting treatment effects are expressed as mean differences, odds ratios (OR), rate ratios (RR), or average hazard ratios (HR). Due to the standardization step, described in Section~\ref{sec:standardization}, the resulting subgroup treatment effect estimates are marginal in the sense that they marginalize over the covariate distribution within the subgroup. The effects are conditional in the sense that they condition on the subgroup.

The linear predictor for subject $i$ has the following form:

\[
  LP_i =  \alpha_0 + \beta_{0}z_i + 
    \underbrace{\alpha_1 s_{i1}+\ldots + \alpha_K s_{iK} + \alpha_{K+1} u_{i1} +\ldots + \alpha_{K+L}u_{iL} }_{\substack{\text{prognostic effects including} \\ \text{main subgroup effects}}}
    + \underbrace{\beta_1 s_{i1} z_i +\ldots + \beta_K s_{iK}z_i}_{\substack{\text{predictive effects: subgroup-by-} \\ \text{treatment interactions}}}  
\]
where $\alpha_0$ and $\beta_0$ are the regression coefficients for the intercept and the main treatment effect. Note that the model for $LP_i$ uses an indicator variable for each level of every categorical covariate. For subgrouping variables to which no shrinkage is applied, some terms may be omitted to avoid overparameterization, as described in Section~\ref{sec:prior}.

Negative binomial models may incorporate an additional offset term to adjust for varying follow-up durations, while Cox regression models exclude the intercept $\alpha_0$.
For the one-way approach, the model can be fully stratified if desired, i.e., assume separate regression coefficients for the additional covariates (i.e., subgroup-by-covariate interactions) and separate variances, overdispersion parameters, or baseline hazards for different levels of the subgrouping variable $x$.

\subsection{Prior distributions}\label{sec:prior}

Regression coefficients for the intercept $\alpha_0$ and the main treatment effect $\beta_0$ are unshrunken and have a flat prior. 
In most cases, including the case study and simulations in this paper, both one-way and global models leave prognostic effects unshrunken and apply a shrinkage prior to all subgroup-by-treatment interaction terms. However, in some applications, it may be advantageous to also shrink some or all prognostic terms, or to leave some subgroup-by-treatment interaction terms in a global model unshrunken. We allow for this flexibility by categorizing the regression coefficients (other than the intercept and the main treatment effect) into three groups: unshrunken terms, shrunken prognostic terms, and shrunken predictive terms:

\begin{itemize}
\item
  \emph{Unshrunken terms}: This group typically consists of all prognostic effects and a non-informative or flat prior is assigned to them. However, when the number of terms is large (relative to the sample size) and regularization is deemed necessary for some or all prognostic effects, they might be moved to the group of shrunken prognostic terms instead. In some applications, there might be a strong a priori rationale for treatment effect heterogeneity for some of the subgroups in a global model, e.g., for those defined by biomarkers directly linked to a drug target. In this case, subgroup-by-treatment interactions for these subgroups might be left unshrunken to reflect this prior knowledge. 
\item
  \emph{Shrunken prognostic terms}: Prognostic terms with shrinkage priors. Often, this group is empty as all prognostic terms are left unshrunken. 
\item
  \emph{Shrunken predictive terms}: Predictive terms with shrinkage priors. For one-way shrinkage models, this group includes the subgroup-by-treatment interactions for the selected subgrouping variable only. For global models, it encompasses all subgroup-by-treatment interactions (with the possible exception of those grouped as unshrunken terms as described above).
\end{itemize}

The design matrix corresponding to unshrunken terms uses dummy encoding, i.e., we omit baseline levels of categorical variables and absorb them in the intercept to avoid overparametrization. For one-way models, this implies that the term $\alpha_1 s_{1i}$ is omitted from the linear predictor. For global models, the terms $\alpha_ks_{ik}$ (for $k \in \{1, l_1+1, l_1+l_2+1, \ldots, \sum_{j=1}^{p-1} l_j+1\}$) corresponding to the baseline levels of each of the $p$ included subgrouping variables, respectively, are omitted. 

For shrunken terms, we aim to treat all levels of the categorical covariates symmetrically. Therefore, we use one hot encoding for shrunken terms, i.e., we include an indicator variable for each level of the categorical covariate accepting an overparameterized model.  

For shrunken terms, we further assume exchangeability for the regression coefficients within the shrunken prognostic and predictive terms separately. In many disease areas, a number of prognostic factors have been established in the clinical literature, whereas firm knowledge of predictive variables is much less common. Thus, an exchangeability assumption across prognostic and predictive terms seems less plausible. The exchangeability assumption for shrunken predictive terms is discussed in more detail in the following subsection.  

In order to guide the definition of reasonable choices for the parameters of hyperpriors for shrunken predictive effects, it is useful to consider what they imply for the prior distribution of coefficients $|\beta_i|$ and of pairwise differences $|\beta_i-\beta_j|$ ($i, j \in\{1,\ldots, K\}$), respectively. These simple checks can help calibrate the prior appropriately and restrict grossly unrealistic areas of the parameter space.

For one-way shrinkage models, the pairwise differences $|\beta_i-\beta_j|$ represent the differences in treatment effects between subgroups $i$ and $j$ conditional on baseline covariates.
If these differences exceed twice the overall treatment effect, this typically indicates a qualitative interaction and substantial heterogeneity.
Since shrinkage priors are symmetric around zero, $\beta_0$ is expected to be similar to the overall treatment effect.
Consequently, values of $|\beta_i|$ ($i\in\{1,\ldots, K\}$) that are larger than the overall treatment effect may also indicate qualitative interactions.

In global models, the coefficients $\beta_i$ are more complex to interpret because $p$ subgroup indicator variables are equal to 1 for every subject.
However, in many settings, heterogeneity may be driven by a single subgrouping variable, such that $\beta_i=0$ for indicator variables corresponding to all other subgroups.
In this scenario, the interpretation used for one-way shrinkage models above also applies to the non-zero coefficients of the heterogeneous subgrouping variable within the global model.

\subsection*{Exchangeability assumptions for shrunken predictive terms in one-way and global shrinkage models} 

As stated above, we assume exchangeability across all shrunken predictive terms in the model. This reflects the belief that prior to observing the data, the labels attached to the interaction terms do not provide information about which interactions are likely to be large and which are likely to be small. 

For one-way shrinkage models, exchangeability is assumed only across the levels of a single subgrouping variable. In many settings, there is no a priori reason to believe that the treatment effect differs systematically from the overall treatment effect for any particular level of the subgrouping variable, making exchangeability a reasonable assumption. The number of levels of a subgrouping variable is typically small (e.g., 2 to 5) and therefore provides limited information to update prior assumptions about heterogeneity. Consequently, one-way shrinkage models are often based on normal priors with a half-normal hyperprior on the standard deviation, so that heterogeneity is characterized by a single parameter. For pragmatic reasons, the same prior specification is typically used across one-way models investigating different subgrouping variables. 

For global shrinkage models, the exchangeability assumption extends across levels of different subgrouping variables and may therefore span variables of very different clinical character (e.g., sex and region). Thus, the model makes a substantially stronger exchangeability assumption than one-way models. In particular, levels belonging to the same subgrouping variable may be more closely related than levels from different variables, which could violate the exchangeability assumption. On the other hand, an advantage of the global approach is that the number of subgroups and corresponding shrunken predictive terms is substantially larger (e.g., 10 to 30 in many trials). Consequently, there is more information available to update prior heterogeneity assumptions, which may reduce sensitivity to the prior specification. Moreover, the larger number of parameters makes it feasible to use more flexible prior distributions. In particular, we adopted a common regularized horseshoe prior to estimate a global sparsity structure across all candidate effect modifiers. Compared with normal priors, the regularized horseshoe allows most shrunken predictive terms to be strongly shrunk toward zero while still permitting a small number of terms to escape shrinkage when supported by the data. Consequently, exchangeability is most naturally interpreted as reflecting uncertainty about which interaction terms correspond to genuine effect modification, rather than an assumption that non-negligible interaction effects should have similar magnitudes. This is compatible with the prior belief that most interaction terms are associated with little or no treatment effect heterogeneity, whereas a small number may be associated with substantial heterogeneity. 

\subsection*{Normal priors with a half-normal hyperprior for the standard deviation}

One-way shrinkage models typically assume that the model coefficients share a common normal prior distribution, with a half-normal ($HN$) prior assigned to their common standard deviation \citep{jones2011,henderson2016,wang2024FDA}:

\[ \beta_1, \ldots, \beta_K \sim N(0,\tau^2) \mbox{ with } \tau \sim HN(\phi). \]

For the choice of the heterogeneity parameter $\phi$, note that this specification implies that the marginal prior distribution for the magnitude of a single coefficient, $|\beta_i|$, has median of $0.37\phi$, with 5\% and 95\% quantiles of $0.01 \phi$ and $2.18 \phi$, respectively.
The corresponding quantities for the pairwise difference $|\beta_i-\beta_j|$ are $0.52 \phi$ (median), $0.02\phi$ (5\% quantile), and $3.09 \phi$ (95\% quantile).
Considerations for selecting the heterogeneity parameter $\phi$ are discussed in 
 Spiegelhalter et al \citep{spiegelhalter2004}, Röver et al \citep{rover2021}, Wang et al \citep{wang2024FDA}, and Bornkamp et al \citep{bornkamp2025}.
Wang et al \citep{wang2024FDA} propose very weakly informative choices of $\phi$ such as $\phi = 0.5$ or $\phi = 1$ for binary, time-to-event, and count outcome where the treatment effect is modeled on the log-scale, or $\phi=\sigma_{plan}$ for continuous outcomes where $\sigma_{plan}$ could be chosen as the standard deviation which informed the sample size calculation of the trial.
However, the low number of subgroups in one-way shrinkage models will typically not be sufficient to inform the posterior for $\tau$ and thus slightly more informative priors could be desirable.
Bornkamp et al \citep{bornkamp2025} consider $\phi = |\delta_{plan}/2|$ or $\phi = |\delta_{plan}|$ where $\delta_{plan}$ represents the target effect size of the trial from the trial protocol.
Choosing $\phi = |\delta_{plan}|$, for example, implies an a priori assumption that pairwise differences in conditional treatment effects between subgroups lie between $0.02\cdot|\delta_{plan}|$ and $3.09\cdot |\delta_{plan}|$ with 90\% probability.
This range seems sufficiently wide to remain conservative, even in settings characterized by substantial heterogeneity.

\subsection*{Regularized horseshoe priors}

The regularized horseshoe prior \citep{piironen2017} is a global-local shrinkage prior of the form
\[ \beta_k|\lambda_k, \tau, c \sim N(0,\tau^2\tilde{\lambda}_k^2)\mbox{ with } \tilde{\lambda}_k^2 = c^2 \lambda_k^2 / (c^2 + \tau^2 \lambda_k^2)\]

with hyper-priors

\[\tau\sim C^+(0,\tau_0^2), \; \lambda_k \sim C^+(0,1)\; (k=1,\ldots,K), \mbox{ and }
c^2 \sim \textrm{Inv-Gamma}(\nu/2,\nu s^2 /2).\]

Global-local shrinkage priors are continuous versions of spike-and-slab priors, i.e., mixture priors of a point mass at zero (the spike) and a continuous distribution which describes the non-zero coefficients (the slab).
They have high concentration around zero to shrink irrelevant coefficients heavily to zero and fat tails to avoid over-shrinking strong signals.
The global shrinkage parameter $\tau$ shrinks all parameters towards zero while the heavy-tailed half-Cauchy prior for $\lambda_k$ allows some parameters to escape this heavy shrinkage.
The slab-component for the regularized horseshoe which governs shrinkage for larger coefficients is a $Student-t_{\nu}(0,s^2)$ distribution \citep{piironen2017}.
As in Wolbers et al \citep{wolbers2025shrinkage}, we choose $s=2$ and $\nu=4$ (the defaults in \texttt{brms}) for binary, time-to-event, and count outcomes where the treatment effect is modeled on the log-scale which implies only very weak regularization for the slab. For continuous outcomes, we choose $s=2\sigma_{plan}$ and $\nu=4$.

Piironen et al \citep{piironen2017} show that prior information about the degree of sparsity in the parameter vector can motivate the choice of the hyperparameter $\tau_0$.
Specifically, they propose the choice $\tau_0= p_0/(K-p_0)\cdot \sigma/\sqrt{n}$ for continuous outcomes where $p_0$ is a prior guess of the expected number of non-zero coefficients.
However, their derivation of the association between the effective number of non-zero coefficients and $\tau_0$ relies on assumptions which are not met for our model, i.e., it assumes a linear model with a regularized horseshoe prior applied to all regression coefficients and uncorrelated covariates with mean 0 and variance 1. Thus, their proposal for the choice of $\tau_0$ should be used with caution in our context.
Instead, one could base the choice of the prior parameters on expected magnitudes of regression coefficients, similarly to the description for normal priors above. In our simulations, we explore the choice $\tau_0 \in \{1,|\delta_{plan}|, |\delta_{plan}|/10\}$ for binary, time-to-event, and count outcomes and $\tau_0 \in \{\sigma_{plan},|\delta_{plan}|, |\delta_{plan}|/10\}$ for continuous endpoints.

The choice $\tau_0=1$ is the default in \texttt{brms} and was also used in Wolbers et al \citep{wolbers2025shrinkage}. For this choice (and $s=2$, $\nu=4$), the marginal prior distribution for the magnitude of a single coefficient, $|\beta_i|$, has median $0.42$, with 5\% and 95\% quantiles of $0.008$ and $3.23$, respectively. This implies only weak regularization.

In the simulation studies for the time-to-event endpoint reported in Section~\ref{simulation-for-a-time-to-event-endpoint} we chose $|\delta_{plan}|=0.3$. For $\tau_0=0.3$, the marginal prior for $|\beta_i|$, has median $0.16$, with 5\% and 95\% quantiles of $0.002$ and $2.30$, respectively, which is still compatible with substantial heterogeneity.
For $\tau_0=0.3/10$, the marginal prior has median $0.02$, with 5\% and 95\% quantiles of $0.0003$ and $0.68$. The marginal prior is now strongly concentrated around 0 but the fat tails still allow values of $|\beta_i|$ more than twice as large as $|\delta_{plan}|$ to remain plausible.

In the simulation studies for the continuous endpoint reported in Section~\ref{simulation-for-a-continuous-endpoint} we chose $\delta_{plan}=0.35$, $\sigma_{plan}=1.20$, a total sample size of $n=500$, and $K=15$ subgroups. 
For the choices $\tau_0 = 0.35$ or $\tau_0 = 0.35/10$, respectively, this implies similar quantiles for the marginal prior of $|\beta_i|/\sigma_{plan}$ as reported for the time-to-event endpoint above. As discussed above, the proposal of Piironen et al \citep{piironen2017} relating $\tau_0$ to the expected number of non-zero coefficients should be used with caution in our context. But if we crudely apply it, then the choice $\tau_0=\delta_{plan}/10$ would imply that we expect approximately 6 of the 15 coefficients to be non-zero.

\subsection*{Other parameters and prior distributions}

Default priors from the R package \texttt{brms} are used for all other parameters.
In particular, this implies flat priors for unshrunken terms.
The Bayesian Cox model for time-to-event outcomes parameterizes the baseline hazard via non-negative M-splines with knots at the quartiles of the observed event times, boundary knots at the minimum and maximum event times (minus or plus the smallest observed difference between event times, respectively), and a Dirichlet prior for the M-spline coefficients.

\subsection{Standardization}\label{sec:standardization}

We use standardization (also referred to as G-computation) to derive marginal treatment effects in subgroups from the outcome model \citep{hernan2023,vanLancker2024}.
In brief, this is implemented as follows.
First, the global model is used to predict potential outcomes for every subject $i$ in the trial twice based on their covariates and assuming that the subject had (hypothetically) been assigned to control or intervention, respectively.
For continuous, binary, or count outcomes, these predictions are obtained as $\hat{Y}_{i,C}=g^{-1}(LP_i(z_i=0))$ and $\hat{Y}_{i,I}=g^{-1}(LP_i(z_i=1))$ where $g^{-1}$ is the inverse link function, and $LP_i(z_i=a)$ is the subject's linear predictor with their treatment assignment modified to control ($a=0$) or intervention ($a=1$), respectively.
For time-to-event outcomes, we similarly predict their survival curves $\hat{S}_{i,C}(t)$ and $\hat{S}_{i,C}(t)$ from the fitted Cox model.
Second, we obtain predicted marginal outcomes or survival curves for a subgroup and control or intervention, respectively, by averaging predictions across subjects from that subgroup.
Third, the treatment effect is obtained as a mean difference (for continuous outcomes), odds ratio (for binary outcomes), or count or rate ratio (for count endpoints) from the predicted marginal outcomes from intervention vs control.
For time-to-event outcomes, we derive the treatment effect estimate from the marginal survival curves as an average hazard ratio as described in Wolbers et al \citep{wolbers2025shrinkage}.
We summarized the treatment effect using the median of the posterior draws as the point estimate, accompanied by a 95\% credible interval bounded by the 2.5\% and 97.5\% quantiles.

\subsection{Software implementation and computational cost}

The methodology is implemented in an R package, \texttt{bonsaiforest2}  \citep{bonsaiforest2}, which builds on \texttt{brms}, a package for Bayesian multi-level regression modelling, and Stan, the computational backend for fitting models using Hamiltonian Monte Carlo sampling \citep{bruckner2017, carpenter2017stan}.

As described in Section~\ref{sec:prior}, the user specifies the treatment, shrunken and unshrunken terms to be included in the outcome model along with the dataset and prior distributions, \texttt{bonsaiforest2} then constructs the necessary dummy variables and calls \texttt{brms} to construct and fit the Stan model.
The resulting posterior distributions are passed back to \texttt{bonsaiforest2} which performs the standardization step and returns the shrunken subgroup estimates.

Generally, we have found these models to be computationally feasible and well behaved with regards to convergence. We fit the models using \texttt{cmdstanr} software and observe a small number of divergent transitions with the default settings, which is to be expected for shrinkage priors and especially the horseshoe prior. This is due to the funnel geometry they create, which is difficult for the sampler to explore. We have found that divergent transitions can be effectively avoided by reducing the step size (\verb|adapt_delta|) with these models.

Based on the simulation in Section~\ref{simulation-for-a-time-to-event-endpoint} for a time-to-event outcome with 10 subgrouping variables and 25 subgroups, we sampled from a dataset with 1000 observations, using 2000 warm-up iterations and 2000 samples from 4 chains in parallel with \verb|adapt_delta|=0.99. The global model with a horseshoe prior took 2 minutes, while the 10 one-way models took 4 minutes. Additionally, both models require a one time compilation time of approximately 20 seconds. These are indicative numbers and will depend on the data and exact sampler settings. 
Absolute computing times vary considerably between computing platforms, sampler settings, prior specification, and the specific data set.

\section{Case study: The GALLIUM trial}\label{sec:gallium}

GALLIUM is a 1:1 randomized trial of obinutuzumab- versus rituximab-based chemotherapy in patients with previously untreated follicular lymphoma.
The primary endpoint is investigator-assessed progression-free survival (PFS).
The trial included 1202 subjects and was powered to detect a target hazard ratio of 0.74 (i.e., $\delta_{plan}=|\log(0.74)|$).

GALLIUM was fully evaluated after the observation of 245 PFS events.
The primary analysis revealed a significant improvement in PFS for the obinutuzumab arm (hazard ratio 0.66, 95\% CI 0.51-0.85, p = 0.001) in the overall population \citep{marcus2017}.
Pre-specified subgroup analyses included those defined by the randomization stratification factors of the Follicular Lymphoma International Prognostic Index (FLIPI) risk group, the chosen chemotherapy backbone, and geographic region.

\begin{figure}

{\centering \includegraphics[width=1\linewidth]{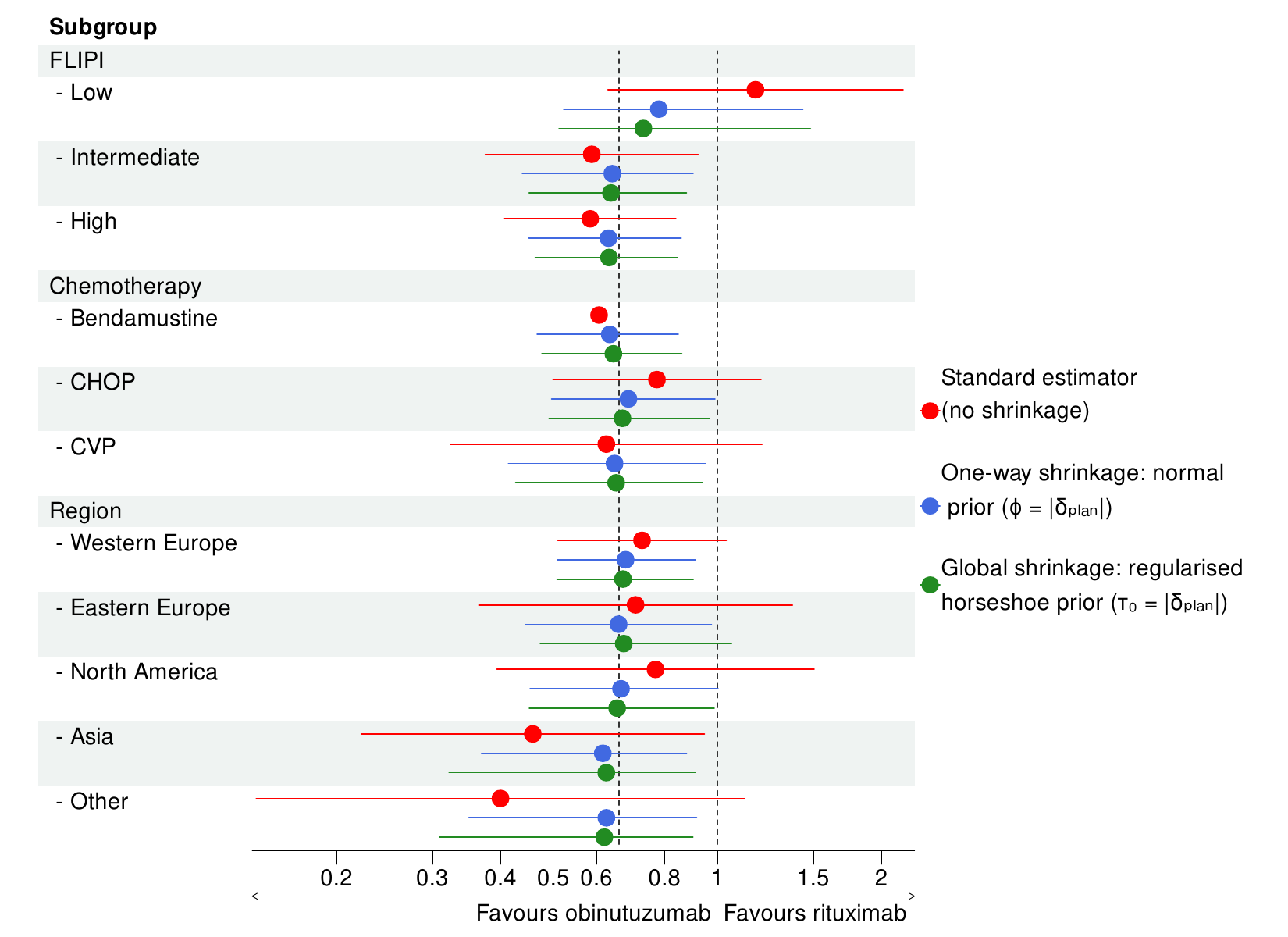} 

}

\caption{Standard and shrinkage treatment effect estimates (average HR with 95\% confidence or credible intervals) in pre-specified subgroups for the GALLIUM trial}\label{fig:fig-gallium-forest}
\end{figure}

Frequentist standard subgroup estimates without shrinkage (as typically shown in forest plots), as well as one-way and global Bayesian shrinkage estimates (with $\phi = |\delta_{plan}|$ and $\tau_0 = |\delta_{plan}|$, respectively) are shown in Figure \ref{fig:fig-gallium-forest}.
The FLIPI low, Asia, and other region subgroups which all contained relatively low event numbers (40, 33, and 18, respectively) were shrunken relevantly towards the overall treatment effect.
In general, the one-way and global shrinkage estimates were similar.

\begin{figure}

{\centering \includegraphics[width=1\linewidth]{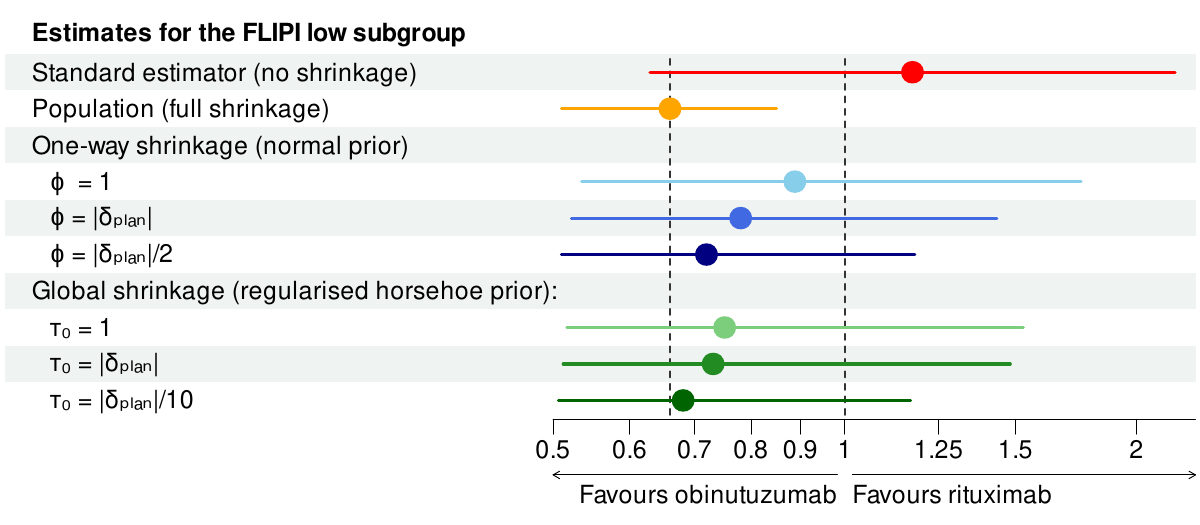} 

}

\caption{Standard and shrinkage treatment effect estimates (average HR with 95\% confidence or credible intervals) in the FLIPI low subgroup of the GALLIUM trial. The standard estimator without shrinkage is in the direction of harm; all shrinkage estimators are in the direction of benefit.}\label{fig:fig-gallium-flipi-low}
\end{figure}

Figure \ref{fig:fig-gallium-flipi-low} shows standard and shrinkage estimates in the FLIPI low subgroup for different choices of prior parameters.
The standard estimator gave a hazard ratio (95\% CI) of 1.17 (0.63-2.19). For the one-way shrinkage model, point estimates were 0.89 ($\phi=1$), 0.78 ($\phi=|\delta_{plan}|$), and 0.72 ($\phi=|\delta_{plan}|/2$).
For the global shrinkage model, point estimates were 0.75 ($\tau_0=1$), 0.73 ($\tau_0=|\delta_{plan}|$), and 0.68 ($\tau_0=|\delta_{plan}|/10$).
The width of the corresponding 95\% credible intervals decreased with increasing informativeness of the prior. The properties of the different shrinkage methods and prior parameter choices are explored further in two simulation studies reported in the next section.

\section{Simulation study}\label{sec:simulation}

We compare one-way and global shrinkage models, utilizing various prior parameter specifications, against standard estimators across two settings.
Simulations for time-to-event endpoints use scenarios from Wolbers et al \citep{wolbers2025shrinkage}, while simulations for continuous endpoints leverage scenarios from Sun et al \citep{sun2024}.

\subsection{Candidate estimators}

One-way shrinkage models included only the subgrouping variable of interest for both prognostic and predictive effects, with flat priors for prognostic effects and shrinkage priors for predictive effects.
Normal shrinkage priors are specified with a half-normal hyperprior for the standard deviation.
Prior parameters are set at $\phi \in \{1, \delta_{plan}, \delta_{plan}/2\}$ for time-to-event endpoints and $\phi \in \{\sigma_{plan}, |\delta_{plan}|, |\delta_{plan}|/2\}$ for continuous endpoints.

Global shrinkage models include all subgrouping variables as prognostic and predictive effects, with flat priors for prognostic effects and shrinkage priors for predictive effects.
We explore regularize horseshoe shrinkage priors with $\tau_0 \in \{1, |\delta_{plan}|, |\delta_{plan}|/10\}$, $s=2$, and $\nu = 4$ for time-to-event and $\tau_0 \in \{\sigma_{plan}, |\delta_{plan}|, |\delta_{plan}|/10\}$,
$s=2\sigma_{plan}$, and $\nu = 4$ for continuous endpoints.
The supplementary materials contain results for global shrinkage models using normal priors with the same parameter specifications as the one-way models.

For comparison, we include the standard subgroup-specific estimator and the population estimator.
These are obtained from a frequentist outcome model with treatment as the sole covariate, applied to subjects from the selected subgroup or the full population, respectively.
They represent the extremes of no shrinkage versus full shrinkage and are typically reported in clinical trial publications.

\subsection{Evaluation criteria}

Performance of different estimators is evaluated across $n_{sim}=1,000$ simulation runs per scenario.
For the overall evaluation, we compare their overall root mean squared error (RMSE), defined as:

\[
\text{RMSE}_{overall} = \sqrt{\frac{1}{n_{sim}K} \sum_{i=1}^{n_{sim}} \sum_{k=1}^K (\hat{\theta}_{ik}-\theta_k)^2}
\]
where $\hat{\theta}_{ik}$ denotes the treatment effect estimate for subgroup $k$ in simulation $i$, and $\theta_k$ denotes the true log-transformed average hazard ratio or mean difference. In our displays, we report the standardized $RMSE_{overall}$ defined as the performance for a specific estimation methods divided by the performance of the standard subgroup-specific estimator without shrinkage as a reference point.
Additionally, we evaluate RMSE, bias, and frequentist coverage of 95\% credible or confidence intervals for each investigated subgroup separately and summarize results as the mean and range across all subgroups included in a scenario.

Scenario 2 for time-to-event outcomes and Scenario 3 for continuous endpoints (described below) represent settings with a substantial overall treatment effect but a lack of efficacy in a specific subgroup.
For this ``null'' subgroup, we report RMSE, bias, and coverage independently.
Furthermore, we assess the selection accuracy by calculating how frequently the most unfavorable subgroup estimate correctly identifies the subgroup lacking efficacy.
Finally, we evaluated the ``bias,'' ``RMSE,'' and ``coverage'' of the 95\% interval for the most extreme subgroup estimate relative to its corresponding true effect.
These metrics are placed in quotation marks to indicate they refer to the performance in a randomly identified subgroup that varies across simulation iterations.

\subsection{Simulation for a time-to-event endpoint}\label{simulation-for-a-time-to-event-endpoint}

\subsubsection*{Data generation}

We re-use simulation scenarios from Wolbers et al \citep{wolbers2025shrinkage} which are motivated by the GALLIUM trial.
In brief, trials with $n=1,000$ and $n_{ev}=247$ events are simulated using a Weibull proportional hazards model. 
This number of events provides 80\% power to detect a hazard ratio of 0.70, i.e., $\delta_{plan}=|log(0.70)|$, at two-sided $\alpha=0.05$.
We include 10 correlated subgrouping variables (25 total subgroups) with sizes ranging from 15\% to 80\% of the total population.
Two variables are prognostic.

We investigate 4 of the 6 scenarios from Wolbers et al \citep{wolbers2025shrinkage} with true average hazard ratios (AHR) evaluated using large simulated datasets:

\begin{itemize}
\setlength{\itemsep}{0pt}\setlength{\parskip}{0pt}
\item
  \emph{Scenario 1 - homogeneous}: No interactions; AHR is approximately 0.66 in all subgroups.
\item
  \emph{Scenario 2 - positive (except one)}: Treatment-by-subgrouping variable interaction for subgrouping variable 4 with three levels resulting in an AHR of 1.00 in subgroup $4a$ (containing 30\% of the population), 0.53 in subgroups $4b$ and $4c$, and 0.68 in all others.
\item
  \emph{Scenario 3 - mild heterogeneity}: Diverse subgroup AHR ranging from 0.70 to 1.17.
\item
  \emph{Scenario 4 - strong heterogeneity}: Diverse subgroup AHR ranging from 0.52 to 1.38.
\end{itemize}

We omitted the "negative except one" scenario from Wolbers et al. \citep{wolbers2025shrinkage} because it provides information similar to that obtained from the "positive except one" scenario. We further omitted the "mis-specified" scenario from Wolbers et al., as model misspecification is already reflected in the continuous-endpoint simulations, all of which are misspecified. Results reported by Pedrera Gomez \citep{pedrera2025} in an unpublished MSc thesis demonstrate that inclusion of these additional scenarios would not materially alter the study conclusions.

Subgroup prevalences and the corresponding true treatment effects underlying each simulation scenario are provided in Supplementary Table S1.
For full details about these scenario we refer to Wolbers et al \citep{wolbers2025shrinkage} and its supplementary materials.

\subsubsection*{Simulation results}

\begin{figure}

{\centering \includegraphics[width=1\linewidth]{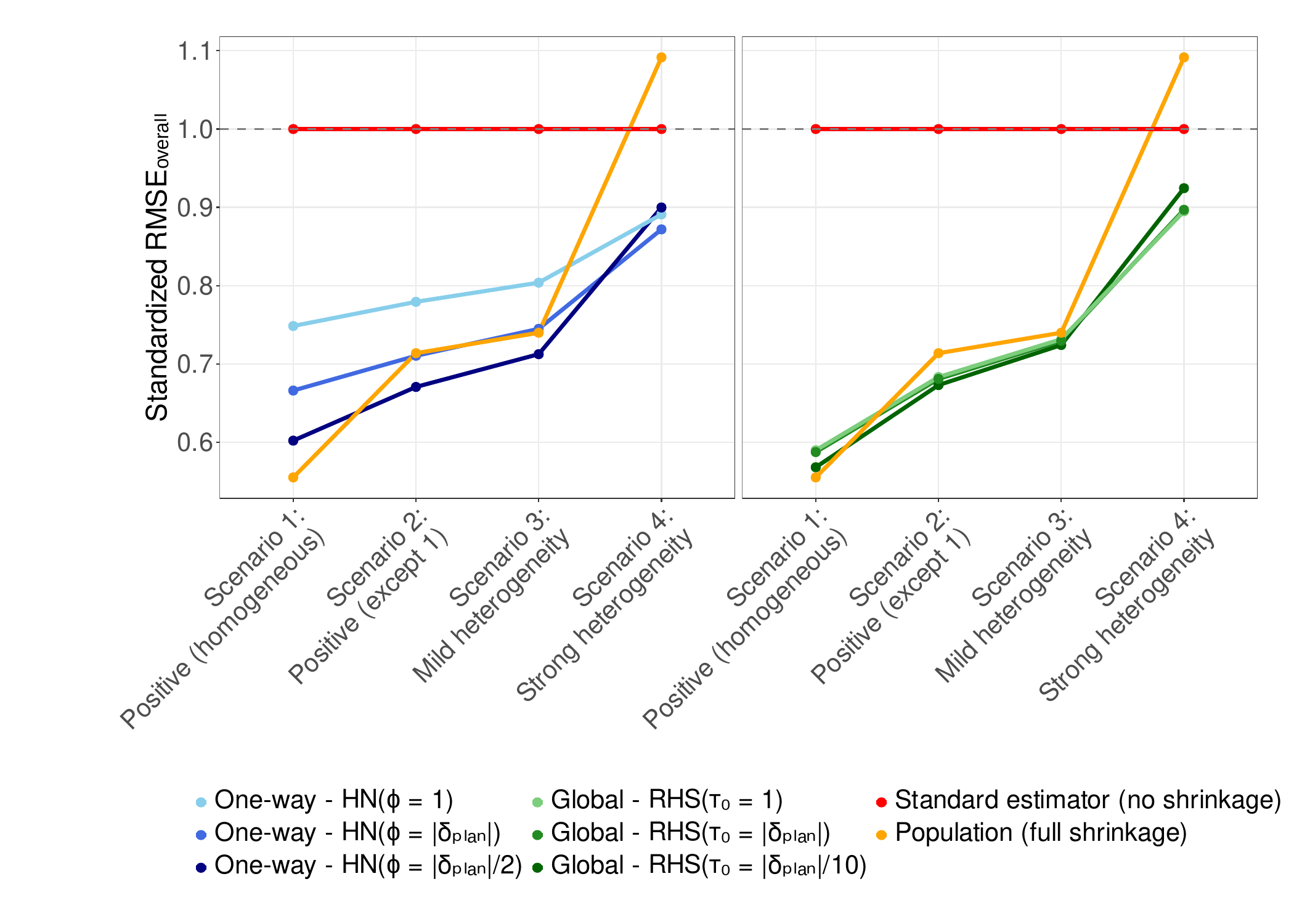} 

}

\caption{Time-to-event endpoint simulation:  Standardized overall root mean squared error by simulation scenario.}\label{fig:fig-tte-stand-RMSE-comparison}
\end{figure}

Standardized $RMSE_{overall}$ for one-way models with a normal prior and global models with a regularized horseshoe prior are shown in Figure \ref{fig:fig-tte-stand-RMSE-comparison}. For the homogeneous scenario 1, standardized $RMSE_{overall}$ values for one-way models decreased from 0.75 to 0.60 as the informativeness of the prior parameter $\phi$ increased. For global models, $RMSE_{overall}$ values ranged from 0.59 to 0.57. For the most heterogeneous scenario 4, standardized $RMSE_{overall}$ values varied from 0.87 to 0.89 for one-way models (with the best performance at $\phi=|\delta_{plan}|$) and from 0.89 to 0.92 for global models (with the best performance at $\tau_0=1$). The population estimator had a lower $RMSE_{overall}$ than the standard estimator across all scenarios except for the most heterogeneous scenario 4. All shrinkage models had lower $RMSE_{overall}$ than the standard estimator without shrinkage across all scenarios. Among one-way models, only the model with the strongest shrinkage ($\phi=\delta_{plan}/2$) had consistently lower $RMSE_{overall}$ across scenarios 2 to 4 than the population estimator. In contrast, all global models were superior to the population estimator across scenarios 2 to 4.

Table \ref{tab:tte-table} summarizes the RMSE, absolute bias, and coverage of nominal 95\% confidence or credible intervals across the 25 subgroups by scenario. The standard estimator without shrinkage had the largest RMSE but minimal bias and accurate coverage across subgroups. For the population estimator, bias was large in scenarios 2 to 4, and the lowest coverage across subgroups in these scenarios ranged from 2\% to 39\%. For all shrinkage estimators, bias was substantially reduced and coverage improved compared to the population estimator. In general, one-way models had slightly lower bias and more favorable coverage than global models in scenarios 2 to 4. For both types of shrinkage models, bias was lower and coverage higher for prior parameters that induced less shrinkage.

\begin{table}[htbp]
\centering
\begin{threeparttable}
\caption{Time-to-event endpoint simulation: Mean (range) of RMSE,  bias, and observed coverage of nominal 95\% confidence or credible intervals across subgroups by simulation scenario.}\label{tab:tte-table}
\begin{tabular}{llrrrr}
\toprule
Metric & Estimator & Scenario 1 & Scenario 2 & Scenario 3 & Scenario 4\\
\midrule
$\rm{RMSE}\cdot 100$ &   &   &   &   &  \\
  & Standard estimator & 23 (15-36) & 23 (15-35) & 22 (14-34) & 23 (14-34)\\
  & Population estimator & 13 (13-13) & 15 (13-41) & 16 (13-31) & 23 (13-53)\\
\addlinespace
  & One-way shrinkage &   &   &   &  \\
  &    $\phi=1$ & 17 (14-23) & 18 (14-23) & 18 (14-26) & 20 (14-34)\\
  &    $\phi = |\delta_{\mathrm{plan}}|$ & 16 (14-19) & 17 (14-24) & 17 (13-25) & 19 (14-35)\\
  &    $\phi = |\delta_{\mathrm{plan}}|/2$ & 14 (13-15) & 15 (13-28) & 16 (13-26) & 20 (13-39)\\
\addlinespace
  & Global shrinkage &   &   &   &  \\
  &    $\tau_0=1$ & 14 (13-15) & 16 (13-28) & 16 (13-27) & 20 (13-37)\\
  &    $\tau_0 = |\delta_{\mathrm{plan}}|$ & 14 (13-15) & 16 (13-29) & 16 (13-27) & 20 (13-38)\\
  &    $\tau_0 = |\delta_{\mathrm{plan}}|/10$ & 13 (13-14) & 15 (13-31) & 16 (13-28) & 20 (13-41)\\
\addlinespace
$|\rm{Bias}|\cdot 100$&   &   &   &   &  \\
  & Standard estimator & 1 (0-2) & 1 (0-2) & 1 (0-2) & 1 (0-2)\\
  & Population estimator & 0 (0-1) & 4 (0-39) & 8 (0-29) & 16 (0-52)\\
\addlinespace
  & One-way shrinkage &   &   &   &  \\
  &    $\phi=1$ & 1 (0-2) & 1 (0-9) & 3 (0-13) & 5 (0-19)\\
  &    $\phi = |\delta_{\mathrm{plan}}|$ & 0 (0-1) & 2 (0-14) & 4 (0-16) & 7 (0-26)\\
  &    $\phi = |\delta_{\mathrm{plan}}|/2$ & 0 (0-1) & 2 (0-22) & 6 (0-20) & 10 (0-35)\\
\addlinespace
  & Global shrinkage &   &   &   &  \\
  &    $\tau_0=1$ & 0 (0-1) & 3 (0-19) & 6 (0-20) & 7 (0-26)\\
  &    $\tau_0 = |\delta_{\mathrm{plan}}|$ & 0 (0-1) & 3 (0-20) & 6 (0-20) & 8 (0-28)\\
  &    $\tau_0 = |\delta_{\mathrm{plan}}|/10$ & 0 (0-1) & 3 (1-24) & 7 (0-23) & 10 (0-33)\\
\addlinespace
Coverage (\%) &   &   &   &   &  \\
  & Standard estimator & 95 (93-96) & 95 (94-96) & 95 (94-97) & 95 (94-97)\\
  & Population estimator & 95 (95-95) & 88 (15-95) & 86 (39-94) & 70 (2-94)\\
\addlinespace
  & One-way shrinkage &   &   &   &  \\
  &    $\phi=1$ & 97 (95-99) & 96 (93-99) & 96 (94-98) & 95 (88-98)\\
  &    $\phi = |\delta_{\mathrm{plan}}|$ & 97 (95-99) & 96 (89-99) & 96 (92-98) & 94 (83-98)\\
  &    $\phi = |\delta_{\mathrm{plan}}|/2$ & 97 (95-99) & 96 (82-99) & 95 (86-98) & 91 (68-98)\\
\addlinespace
  & Global shrinkage &   &   &   &  \\
  &    $\tau_0=1$ & 96 (94-98) & 95 (81-99) & 94 (84-98) & 93 (81-97)\\
  &    $\tau_0 = |\delta_{\mathrm{plan}}|$ & 97 (95-98) & 95 (81-98) & 94 (83-98) & 92 (79-97)\\
  &    $\tau_0 = |\delta_{\mathrm{plan}}|/10$ & 96 (95-98) & 94 (73-98) & 92 (76-98) & 90 (71-97)\\
\bottomrule
\end{tabular}
\begin{tablenotes}[flushleft]
\footnotesize
\item Each scenario includes 25 subgroups. RMSE and bias are multiplied by 100 to improve readability. Mean (maximum) Monte Carlo standard errors across estimators, scenarios, and subgroups are 0.5 (1.0) for $\rm{RMSE}\cdot 100$, 0.5 (1.1) for $\rm{Bias}\cdot 100$, and 0.7\% (1.6\%) for coverage.
\end{tablenotes}
\end{threeparttable}
\end{table}

Scenario 2 includes one subgroup (subgroup $4a$) for which the treatment is inefficacious (true AHR = 1). Table \ref{tab:tte-table-worst} shows the performance of estimators in subgroup $4a$, as well as in the subgroup with the worst predicted treatment effect for each model and simulation run. For subgroup $4a$, the one-way model with $\phi=1$ had the lowest RMSE and bias, and the best coverage among shrinkage estimators. Other shrinkage estimators had larger RMSE and bias; coverage ranged from 82\% to 93\% for one-way models and from 73\% to 81\% for global models. Outside of simulation studies, the true subgroup treatment effects are unknown, but researchers are often interested in the subgroup with the worst observed treatment effect. Shrinkage estimators had higher accuracy in identifying subgroup $4a$ as the worst subgroup (with accuracy ranging from 63\% to 67\%) than the standard estimator, which had an accuracy of 51\%. Bias, RMSE, and coverage in that randomly identified subgroup were also substantially worse for the standard estimator compared to all shrinkage estimators. Among shrinkage estimators, bias and RMSE for the worst observed subgroup were largest for the one-way model with $\phi=1$.

\begin{table}[htbp]
\centering
\begin{threeparttable}
\caption{Time-to-event endpoint simulation (scenario 2): Performance of estimators in subgroup 4a where the drug does not work (true $\log(AHR)=0$) and in the subgroup with the worst predicted treatment effect.}\label{tab:tte-table-worst}
\begin{tabular}{lccccccc}
\toprule
\multicolumn{1}{c}{ } & \multicolumn{3}{c}{Subgroup 4a} & \multicolumn{4}{c}{Worst Predicted Subgroup} \\
\cmidrule(l{3pt}r{3pt}){2-4} \cmidrule(l{3pt}r{3pt}){5-8}
\makecell[c]{Estimator\\~} & \makecell[c]{RMSE\\~} & \makecell[c]{Bias\\~} & \makecell[c]{Coverage\\(\%)} & \makecell[c]{Accuracy\\(\%)} & \makecell[c]{RMSE\\~} & \makecell[c]{Bias\\~} & \makecell[c]{Coverage\\(\%)}\\
\midrule
Standard estimator & 0.20 & 0.00 & 95 & 51 & 0.40 & 0.29 & 81\\
Population estimator & 0.41 & -0.39 & 15 &  &  &  & \\
\addlinespace
One-way shrinkage &  &  &  &  &  &  & \\
$\phi=1$ & 0.22 & -0.09 & 93 & 63 & 0.27 & 0.11 & 91\\
$\phi = |\delta_{\mathrm{plan}}|$ & 0.24 & -0.14 & 89 & 65 & 0.24 & 0.03 & 92\\
$\phi = |\delta_{\mathrm{plan}}|/2$ & 0.28 & -0.22 & 82 & 67 & 0.23 & -0.06 & 93\\
\addlinespace
Global shrinkage &  &  &  &  &  &  & \\
$\tau_0=1$ & 0.28 & -0.19 & 81 & 65 & 0.24 & -0.03 & 92\\
$\tau_0 = |\delta_{\mathrm{plan}}|$ & 0.29 & -0.20 & 81 & 65 & 0.25 & -0.04 & 91\\
$\tau_0 = |\delta_{\mathrm{plan}}|/10$ & 0.31 & -0.24 & 73 & 65 & 0.26 & -0.08 & 89\\
\bottomrule
\end{tabular}
\begin{tablenotes}[flushleft]
\footnotesize
\item Accuracy is defined as the proportion of simulation runs in which subgroup 4a was correctly identified as having the worst predicted treatment effect. Performance for the worst predicted subgroup is evaluated by comparing the estimate to the true value within the identified subgroup.
\end{tablenotes}
\end{threeparttable}
\end{table}

In addition to global models with a regularized horseshoe prior, we also investigated global models with a normal prior and report these results in the supplementary materials. In general, their performance (reported in Table S3) was comparable to that of models with a regularized horseshoe prior (Table \ref{tab:tte-table}). However, their performance for the worst subgroup in scenario 2 was worse, showing poorer accuracy in identifying the worst subgroup (62\% versus 65\% for the regularized horseshoe prior) and lower coverage for both subgroup $4a$ and the treatment effect in the worst identified subgroup (Tables S4 and \ref{tab:tte-table-worst}).

\subsection{Simulation for a continuous endpoint}\label{simulation-for-a-continuous-endpoint}

\subsubsection*{Data generation}

Sun et al~proposed data-generating mechanisms to compare algorithms for characterizing treatment effect heterogeneity in randomized trials and implemented them in the R package \texttt{benchtm} \citep{sun2024,benchtm}.
For our simulations, we adapt their Scenario 1 for continuous outcomes with increasing degrees of heterogeneity by modifying the simulation parameters $\beta_0$ to guarantee a power of 90\% for the overall treatment comparison (compared to 50\% in Sun et al \citep{sun2024}).

Baseline covariates are generated as synthetic data mimicking the joint distribution of a pool of real phase III trials for a pharmaceutical compound in an inflammatory disease as described in Sun et al \citep{sun2024}.
Amongst the generated covariates we chose the first few categorical covariates plus all covariates which had a direct prognostic or predictive effect for any of the scenarios proposed in Sun et al \citep{sun2024}. We avoided covariates which had low prevalences for some levels because this would deteriorate the performance of the subgroup-specific estimator without shrinkage. Specifically, we use the following six covariates as subgrouping variables describing 15 subgroups of interest: The binary covariates $X_1, X_2, X_4$ and $X_8$, dichotomized versions of $X_{11}$ and $X_{17}$ with cut-points at their median, and a categorized version of $X_{14}$ with cut-points defined by the lower quartile and the median.
Subgroups defined by all subgrouping variables have prevalences of at least 25\%.

Continuous outcomes are generated according to the following model:
\[ Y = 2.30\cdot(0.5I(X_1=\text{``Y''}) + X_{11}) + Z (\beta_0 + \beta_1\Phi(20(X_{11}-0.5))) + \epsilon 
  \mbox{ with } \epsilon \sim N(0,1), \]
where $Z\in \{0,1\}$ denotes treatment assignment and $\Phi(.)$ is the standard normal cumulative distribution function.
$X_1$ and $X_{11}$ are prognostic effects, the parameter $\beta_0$ controls the conditional treatment effect at $X_{11}=0.5$, and $\beta_1$ controls the predictive effect of $X_{11}$.

Randomized clinical trials with a 1:1 randomization ratio and a total sample size of $n=500$ subjects are simulated from three scenarios:

\begin{itemize}
\setlength{\itemsep}{0pt}\setlength{\parskip}{0pt}
\item
  \emph{Scenario 1 - homogeneous} ($\beta_0=0.35, \beta_1=0$): The population treatment effect is 0.35 and the within group standard deviation is 1.21.
\item
  \emph{Scenario 2 - moderate heterogeneity} ($\beta_0=0.19, \beta_1=0.38$): $\beta_1$ is chosen to achieve a power of 80\% for the interaction test of $H_0:\beta_1=0$ versus $H_1:\beta_1\neq 0$ (as described in Sun et al \citep{sun2024}). The population treatment effect is 0.36 and the within group standard deviation is 1.24. Treatment effects in subgroups defined by the dichotomized variable $X_{11}$ are 0.21 and 0.52. Treatment effects in all other subgroups range from 0.31 to 0.39.
\item
  \emph{Scenario 3 - strong heterogeneity} ($\beta_0=0.04, \beta_1=0.77$): $\beta_1$ is twice as large as for the moderate heterogeneity scenario. The population treatment effect is 0.37 and the within group standard deviation is 1.28. Treatment effects in subgroups defined by the dichotomized variable $X_{11}$ are 0.06 (i.e., close to ``null'') and 0.69. Treatment effects in all other subgroups range from 0.27 to 0.44.
\end{itemize}

The simulation truth is derived from a synthetic population of 50,000 subjects.
Covariates of trial participants are sampled from this population. Subgroup prevalences and the corresponding true treatment effects underlying each simulation scenario are provided in Supplementary Table S2.

Parameters for shrinkage distributions use $\delta_{plan}=0.35$ and $\sigma_{plan}=1.20$ for all scenarios.

\subsubsection*{Simulation results}

\begin{figure}

{\centering \includegraphics[width=1\linewidth]{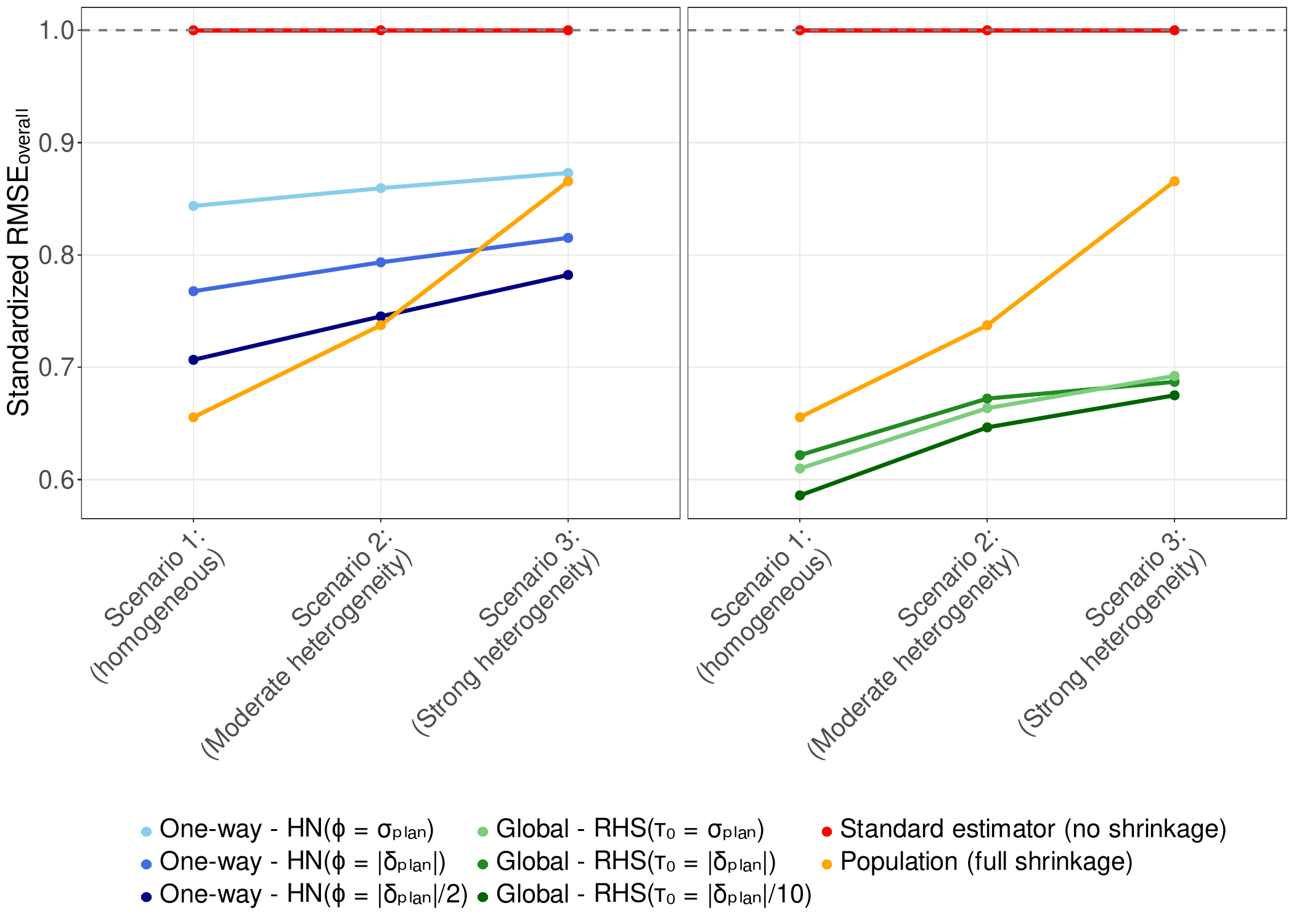} 

}

\caption{Continuous endpoint simulation:  Standardized overall root mean squared error by simulation scenario.}\label{fig:fig-cont-stand-RMSE-comparison}
\end{figure}

Standardized $RMSE_{overall}$ values for one-way models with a normal prior and global models with a regularized horseshoe prior are shown in Figure \ref{fig:fig-cont-stand-RMSE-comparison}. The population estimator and all shrinkage estimators had lower $RMSE_{overall}$ values across all scenarios than the standard estimator without shrinkage. As in the simulation with a time-to-event endpoint, the choice of the prior parameter impacted the results for one-way models more strongly than for global models. In particular, standardized $RMSE_{overall}$ values for the one-way model with $\phi=1$ were between 0.09 to 0.14 higher than for the choice $\phi=|\delta_{plan}|/2$ across the three scenarios. Notably, global models with a regularized horseshoe prior had lower $RMSE_{overall}$ values than all other estimators across all scenarios, including the population estimator in the homogeneous scenario 1. Further investigation showed that this occurs because global shrinkage adjusts for all subgrouping variables as prognostic factors in the models, hence benefiting from a covariate adjustment effect \citep{vanLancker2024}, whereas the population estimator does not adjust for any covariates and the one-way models only adjust for the subgrouping variable included in the respective model. In the supplementary materials, we report results for covariate-adjusted population estimators and estimates from one-way models with adjustment for all subgrouping variables as unshrunken prognostic effects (Figure S1). After covariate adjustment, the population estimator had the lowest $RMSE_{overall}$ value for the homogeneous scenario 1, and the performance of the one-way model with the most informative prior, $\phi=|\delta_{plan}|/2$, became more similar to the results for the global model.

Table \ref{tab:cont-table} summarizes the RMSE, absolute bias, and coverage of nominal 95\% confidence or credible intervals across the 15 subgroups by scenario. The standard estimator without shrinkage had the largest RMSE but minimal bias and accurate coverage across subgroups. The population estimator had a lower RMSE than the standard estimator but the largest bias and the poorest coverage among all estimators in scenarios 2 and 3.
In the comparison between one-way and global models, the RMSE was lowest for the global models, whereas maximum bias and minimum credible interval coverage tended to be more favorable for one-way models. One-way models with covariate adjustment had improved RMSE compared to those without adjustment and, for the the most informative prior, $\phi=|\delta_{plan}|/2$, RMSE was comparable to global models (Supplementary Table S7).

\begin{table}[htbp]
\centering
\begin{threeparttable}
\caption{Continuous endpoint simulation: Mean (range) of RMSE,  bias, and observed coverage of nominal 95\% confidence or credible intervals across subgroups by simulation scenario.}\label{tab:cont-table}
\begin{tabular}{llrrr}
\toprule
Metric & Estimator & Scenario 1 & Scenario 2 & Scenario 3\\
\midrule
$\rm{RMSE}\cdot 100$ &    &    &    &  \\
  & Standard estimator & 16 (12-23) & 16 (13-23) & 17 (14-24)\\
  & Population estimator & 10 (10-10) & 12 (11-19) & 15 (12-34)\\
\addlinespace
  & One-way shrinkage &    &    &  \\
  &    $\phi = \sigma_{\mathrm{plan}}$ & 13 (11-16) & 14 (12-17) & 15 (13-18)\\
  &    $\phi = |\delta_{\mathrm{plan}}|$ & 12 (11-14) & 13 (11-15) & 14 (12-17)\\
  &    $\phi = |\delta_{\mathrm{plan}}|/2$ & 11 (10-12) & 12 (11-15) & 13 (12-19)\\
\addlinespace
  & Global shrinkage &    &    &  \\
  &    $\tau_0 = \sigma_{\mathrm{plan}}$ & 10 (9-10) & 11 (10-15) & 12 (10-18)\\
  &    $\tau_0 = |\delta_{\mathrm{plan}}|$ & 10 (9-11) & 11 (10-15) & 12 (10-17)\\
  &    $\tau_0 = |\delta_{\mathrm{plan}}|/10$ & 9 (9-10) & 11 (10-16) & 12 (10-20)\\
\addlinespace
$|\rm{Bias}|\cdot 100$ &    &    &    &  \\
  & Standard estimator & 0 (0-1) & 0 (0-1) & 0 (0-1)\\
  & Population estimator & 0 (0-0) & 3 (0-16) & 6 (0-32)\\
\addlinespace
  & One-way shrinkage &    &    &  \\
  &    $\phi = \sigma_{\mathrm{plan}}$ & 0 (0-1) & 1 (0-4) & 1 (0-3)\\
  &    $\phi = |\delta_{\mathrm{plan}}|$ & 0 (0-1) & 1 (0-6) & 2 (0-7)\\
  &    $\phi = |\delta_{\mathrm{plan}}|/2$ & 0 (0-1) & 2 (0-9) & 3 (0-12)\\
\addlinespace
  & Global shrinkage &    &    &  \\
  &    $\tau_0 = \sigma_{\mathrm{plan}}$ & 0 (0-0) & 2 (0-9) & 2 (0-8)\\
  &    $\tau_0 = |\delta_{\mathrm{plan}}|$ & 0 (0-0) & 1 (0-8) & 1 (0-6)\\
  &    $\tau_0 = |\delta_{\mathrm{plan}}|/10$ & 0 (0-0) & 2 (0-11) & 2 (0-11)\\
\addlinespace
Coverage (\%) &    &    &    &  \\
  & Standard estimator & 95 (94-96) & 95 (93-96) & 94 (93-95)\\
  & Population estimator & 96 (96-96) & 91 (70-95) & 84 (19-95)\\
\addlinespace
  & One-way shrinkage &    &    &  \\
  &    $\phi = \sigma_{\mathrm{plan}}$ & 97 (96-98) & 96 (94-98) & 96 (94-98)\\
  &    $\phi = |\delta_{\mathrm{plan}}|$ & 97 (96-98) & 96 (93-98) & 96 (92-98)\\
  &    $\phi = |\delta_{\mathrm{plan}}|/2$ & 98 (96-98) & 96 (91-98) & 95 (88-98)\\
\addlinespace
  & Global shrinkage &    &    &  \\
  &    $\tau_0 = \sigma_{\mathrm{plan}}$ & 97 (96-98) & 96 (89-98) & 95 (88-97)\\
  &    $\tau_0 = |\delta_{\mathrm{plan}}|$ & 97 (96-98) & 96 (90-98) & 95 (91-97)\\
  &    $\tau_0 = |\delta_{\mathrm{plan}}|/10$ & 97 (96-98) & 95 (85-97) & 94 (85-96)\\
\bottomrule
\end{tabular}
\begin{tablenotes}[flushleft]
\footnotesize
\item Each scenario includes 15 subgroups. RMSE and bias are multiplied by 100 to improve readability. Mean (maximum) Monte Carlo standard errors across estimators, scenarios, and subgroups are 0.3 (0.6) for $\rm{RMSE}\cdot 100$, 0.4 (0.7) for $\rm{Bias}\cdot 100$, and 0.6\% (1.5\%) for coverage.
\end{tablenotes}
\end{threeparttable}
\end{table}

Scenario 3 includes one subgroup (subgroup $11a$) for which the true treatment effect was substantially worse than in all other subgroups (0.06 versus $\geq 0.27$). Table \ref{tab:cont-table-worst} shows the performance of estimators in subgroup $11a$, as well as in the subgroup with the worst predicted treatment effect for each model and simulation run. The bias of the population estimator was 0.30, compared to values between 0.03 and 0.11 for all shrinkage estimators. Global shrinkage estimators were most successful in identifying the subgroup with the worst treatment effect (accuracy of 89\% to 90\% for different choices of $\tau_0$) compared to one-way estimators (77\% to 79\% for different choices of $\phi$) and the standard estimator without shrinkage (70\%). For covariate-adjusted one-way shrinkage models, accuracy was comparable to the global models (87\% to 89\% for different choices of $\phi$, Supplemantary Table S8).

\begin{table}[htbp]
\centering
\begin{threeparttable}
\caption{Continuous endpoint simulation (scenario 3): Performance of estimators in subgroup 11a which has the worst true efficacy (true treatment effect = 0.06) and in the subgroup with the worst predicted treatment effect.}\label{tab:cont-table-worst}
\begin{tabular}{lccccccc}
\toprule
\multicolumn{1}{c}{ } & \multicolumn{3}{c}{Subgroup 11a} & \multicolumn{4}{c}{Worst Predicted Subgroup} \\
\cmidrule(l{3pt}r{3pt}){2-4} \cmidrule(l{3pt}r{3pt}){5-8}
\makecell[c]{Estimator\\~} & \makecell[c]{RMSE\\~} & \makecell[c]{Bias\\~} & \makecell[c]{Coverage\\(\%)} & \makecell[c]{Accuracy\\(\%)} & \makecell[c]{RMSE\\~} & \makecell[c]{Bias\\~} & \makecell[c]{Coverage\\(\%)}\\
\midrule
Standard estimator & 0.15 & -0.00 & 94 & 70 & 0.24 & -0.13 & 85\\
Population estimator & 0.33 & 0.30 & 24 &  &  &  & \\
\addlinespace
One-way shrinkage &  &  &  &  &  &  & \\
$\phi= \sigma_{\mathrm{plan}}$ & 0.16 & 0.03 & 94 & 77 & 0.19 & -0.06 & 91\\
$\phi = |\delta_{\mathrm{plan}}|$ & 0.16 & 0.06 & 92 & 79 & 0.18 & -0.02 & 93\\
$\phi = |\delta_{\mathrm{plan}}|/2$ & 0.18 & 0.11 & 88 & 79 & 0.17 & 0.03 & 93\\
\addlinespace
Global shrinkage &  &  &  &  &  &  & \\
$\tau_0=\sigma_{\mathrm{plan}}$ & 0.17 & 0.07 & 89 & 90 & 0.16 & 0.04 & 92\\
$\tau_0 = |\delta_{\mathrm{plan}}|$ & 0.16 & 0.05 & 91 & 89 & 0.16 & 0.01 & 93\\
$\tau_0 = |\delta_{\mathrm{plan}}|/10$ & 0.19 & 0.10 & 86 & 90 & 0.17 & 0.06 & 90\\
\bottomrule
\end{tabular}
\begin{tablenotes}[flushleft]
\footnotesize
\item Accuracy is defined as the proportion of simulation runs in which subgroup 11a was correctly identified as having the worst predicted treatment effect. Performance for the worst predicted subgroup is evaluated by comparing the estimate to the true value within the identified subgroup.
\end{tablenotes}
\end{threeparttable}
\end{table}

Global models with a normal prior had slightly worse performance compared to global models with a regularized horseshoe prior. In particular, the minimum coverage and the accuracy in identifying the subgroup with the worst treatment effect were lower (Tables S5 and S6).

\section{Discussion}\label{sec:discussion}

We presented a unified modeling framework for one-way and global shrinkage models to obtain treatment effect estimates in subgroups. The framework contains a model-fitting and a standardization step. While standardization was not included in earlier presentations of one-way models (e.g., Wang et al \citep{wang2024FDA}), it is essential to obtain treatment effect estimates in subgroups which are not conditional on the included prognostic covariates for non-collapsible treatment effect measures \citep{daniel2021}. Both one-way and global shrinkage models for continuous, binary, count, and time-to-event endpoints are implemented in our R package \texttt{bonsaiforest2}  \citep{bonsaiforest2}.

There are important structural differences between one-way and global shrinkage models. A one-way model is particularly well suited when the objective is to investigate treatment effect heterogeneity with respect to a single subgrouping variable of interest, for example potential regional differences. One-way shrinkage models without covariate adjustment can also be fitted to summary-level data \citep{wang2024FDA}. Furthermore, when there are concerns about model misspecification, the one-way model can be fully stratified, allowing separate covariate effects, variances, or baseline hazards across levels of the subgrouping variable. Investigating heterogeneity with respect to multiple subgrouping variables using the one-way approach requires fitting separate models for each variable. These models may not be mutually compatible. For example, if treatment effect heterogeneity truly exists with respect to only one subgrouping variable, then the one-way models fitted for the remaining variables will necessarily be misspecified because they typically omit these true sources of heterogeneity in their default parameterizations.

The global shrinkage approach addresses this limitation by fitting a single regression model that allows for treatment effect heterogeneity across all considered subgrouping variables simultaneously. Beyond providing subgroup-specific treatment effect estimates, inspection of the fitted model may also yield insights into which variables are driving heterogeneity. The one way models cannot always distinguish true drivers of treatment effect heterogeneity with correlates of these drivers if they are only included in separate models. Due to their complexity, global models cannot be fitted to summary-level data and require access to patient-level data. Stratification with respect to one or more subgrouping variables is also feasible for global models, provided that the sample sizes within strata defined by combinations of these variables are not too small, but full stratification across all subgrouping variables is typically not feasible. 
As discussed in Section~\ref{sec:prior}, our implementation of the global model relies on an exchangeability assumption across subgrouping variables. This assumption is stronger, and arguably less plausible, than the exchangeability assumptions underlying one-way models. However, it permits a more flexible prior specification and may reduce sensitivity to the choice of prior parameters. An alternative class of global shrinkage models could assume exchangeability only within the levels of each subgrouping variable. Such an approach would likely strengthen the plausibility of the exchangeability assumption, but at the cost of introducing additional prior parameters and potentially increasing sensitivity to their specification. We have not explored such models here, but they represent an interesting direction for future research.

To our knowledge, this is the first simulation study for treatment effect estimation in subgroups that compares one-way and global shrinkage models, the standard estimator without shrinkage, and the population estimator. We included two published simulation approaches: one for a time-to-event endpoint and the second for a continuous endpoint. For the time-to-event endpoint, the simulation was inspired by our case study of a clinical trial in follicular lymphoma. For the continuous endpoint, covariates were generated based on a synthetic version of the joint distribution of covariates observed in clinical trials of an inflammatory disease, and the data-generating model included non-linear terms (i.e., the data-generating and the analysis models were not identical).

Across all simulations, all investigated shrinkage estimators had a lower overall root mean squared error ($RMSE$) than the standard estimator without shrinkage. Reductions in overall $RMSE$ ranged from 8\% to 43\% depending on the shrinkage estimator and scenario. In terms of overall $RMSE$, the population estimator was also superior to the standard estimator in all simulations except for one scenario with large treatment effect heterogeneity, confirming earlier findings \citep{wolbers2025shrinkage}. As expected, the substantial reduction in mean squared error provided by the shrinkage estimators came at the cost of introducing some bias into the treatment effect estimates across repeated samples. Consequently, the frequentist coverage of nominal 95\% credible intervals was below the nominal level for some subgroups. This behavior is unsurprising because Bayesian credible intervals are not constructed to guarantee nominal frequentist coverage. Nevertheless, coverage remains a useful performance measure, as it quantifies the impact of shrinkage on interval estimation.  

Two simulation scenarios considered settings in which treatment was efficacious in the overall population but a subgroup existed for which treatment efficacy was minimal or absent. Interestingly, all shrinkage methods were more successful than the standard estimator in identifying this subgroup. In the randomly selected subgroup with the lowest estimated treatment effect, shrinkage estimators also yielded lower bias and mean squared error. Notably, credible intervals based on the shrinkage methods achieved substantially higher and more accurate frequentist coverage than those obtained using the standard estimator.

In general, the differences between individual shrinkage estimators were substantially smaller than the differences between shrinkage-based approaches and the standard estimator without shrinkage. Among the shrinkage estimators, no clear overall winner emerged, either between one-way and global models or among the different prior specifications considered. Beyond the structural differences between one-way and global models discussed above, our simulation results suggest that the most appropriate shrinkage approach depends on the characteristics of the specific application. 

In general, global models tended to have a lower overall $RMSE$. Comparable performance was observed for one-way models only when the most informative prior, $\phi = |\delta_{plan}|/2$, was used together with appropriate covariate adjustment. However, global models were also associated with larger bias and undercoverage of credible intervals in subgroups associated with substantial treatment effect heterogeneity. Similarly, more informative hyperpriors for the shrinkage parameters generally resulted in lower overall $RMSE$ values, except in settings with substantial treatment effect heterogeneity, but also tended to increase bias and undercoverage. The impact of the hyperprior specification on overall $RMSE$ was more pronounced for one-way than for global shrinkage models. One possible explanation is that global models borrow information across a larger number of subgroups, such that the data play a greater role in updating the prior and the resulting estimates are therefore less sensitive to the choice of hyperprior. 

Global models inherently incorporate covariate adjustment through the inclusion of prognostic effects associated with the subgrouping variables. Equivalent covariate adjustment can also be applied in one-way models and, as demonstrated by our simulations, can lead to substantial reductions in mean squared error. More generally, adjusting for important prognostic factors is expected to improve the precision of both one-way and global shrinkage estimators.

We also explored global models with a normal prior, but they tended to perform less well, particularly in settings where substantial treatment effect heterogeneity was driven by a single subgrouping variable. In such settings, the regularized horseshoe prior, which strongly shrinks small signals towards zero while allowing a limited number of larger effects to escape substantial shrinkage, appeared better suited than a normal prior with rapidly decaying tails. Normal priors may, however, be more appropriate to global shrinkage models that assume exchangeability only within the levels of each subgrouping variable. 

Bornkamp et al \citep{bornkamp2025} proposed to base the choice of the parameters of the hyperprior on the anticipated treatment effect size $\delta_{plan}$ and, in the case of continuous data, the expected standard deviation $\sigma_{plan}$. These quantities are essential for the design of a clinical trial and should therefore always be available. Based on our simulations, the choice of $\phi = |\delta_{plan}|$ for one-way models with a normal prior, and the choice of $\tau_0 = |\delta_{plan}|$, $s = 2\sigma_{plan}$ for continuous and $s=2$ for other endpoints, and $\nu = 4$ for global models with a regularized horseshoe prior appear to be sensible defaults. We acknowledge that substantial discrepancies between the planned and observed treatment effect or standard deviation may render a prior anchored in the trial planning assumptions either overly restrictive or insufficiently informative. More generally, and especially in such settings, we recommend conducting sensitivity analyses with respect to the prior assumptions, as illustrated in our case study (Figure 2).  

Our simulation study focused on time-to-event and continuous endpoints. For each endpoint type, we considered a substantially different data-generating mechanism to increase the breadth of scenarios covered. Although binary and count endpoints were not investigated, we expect the main conclusions to extend to these settings. Supporting evidence is provided by simulation results for all four endpoint types reported by Pedrera Gomez \citep{pedrera2025} in an unpublished MSc thesis, which suggest broadly similar performance of shrinkage methods across endpoint types. Nevertheless, additional simulation studies are warranted to further evaluate the performance of shrinkage methods across a wider range of endpoint types and data-generating mechanisms. 
Moreover, we encourage further empirical benchmarking studies for shrinkage estimators in settings with twin concurrent phase 3 trials \citep{bornkamp2025}.

Publications on Bayesian shrinkage priors often focus on regression models with continuous, standardized, and uncorrelated covariates, with the same shrinkage prior applied to all regression coefficients. In contrast, in our setting, covariates are categorical and shrinkage is only applied to a subset of regression coefficients. More research is needed to motivate hyperprior choices for common shrinkage priors, such as the regularized horseshoe or the R2-D2 prior targeted to our setting, and to develop alternative priors for categorical covariates \citep{zhang2022bayesian}. 

In conclusion, we recommend that shrinkage estimators for treatment effects in subgroups be more routinely implemented because they typically provide more realistic estimates of the treatment effect in subgroups than the standard estimator without shrinkage or the population estimator with full shrinkage. In forest plots specifically, shrinkage estimators are a natural complement to the standard estimator and should more frequently be displayed.

\section*{Software and simulation code}

All shrinkage methods are implemented in the R package \texttt{bonsaiforest2} \citep{bonsaiforest2} (developed at \url{ https://github.com/openpharma/bonsaiforest2/}) which comes with detailed vignettes and documentation.
The simulation code is available in the \texttt{simulations} subfolder of this repository.

\section*{Data availability statement}

Qualified researchers may request access to individual patient level data of the GALLIUM trial \citep{marcus2017} through the clinical study data request platform \url{https://vivli.org}.
Further details on Roche's criteria for eligible studies are available here: \url{https://vivli.org/members/ourmembers}.
For further details on Roche's Global Policy on the Sharing of Clinical Information and how to request access to related clinical study documents, see here: \url{https://www.roche.com/research_and_development/who_we_are_how_we_work/clinical_trials/our_commitment_to_data_sharing.htm}.

\section*{Acknowledgements}

The authors thank Björn Bornkamp and Sebastian Weber from Novartis for helpful discussions which improved this manuscript.

\section*{Funding statement}

The majority of this research was conducted while all authors were employed by Roche. Additional work on the manuscript was completed while Marcel Wolbers and Isaac Gravestock were employed by Bayer BCC AG and Miriam Pedrera Gómez was employed by Indivi. No external funding was received for this work.

\bibliography{SubgroupRefs.bib}
\renewcommand{\bibliography}[1]{}

\clearpage
\appendix

\begingroup
\changefontsizes{11pt}
\section{Appendix: Supplementary Tables and Figures}
\setcounter{table}{0}
\renewcommand{\thetable}{S\arabic{table}}
\setcounter{figure}{0}
\renewcommand{\thefigure}{S\arabic{figure}}

\subsection{Simulation truth for time-to-event and continuous endpoint simulations}

Table \ref{tab:tte-truthtable} summarizes the true treatment effects in the overall population and across all subgroups for the time-to-event endpoint simulations. Full details of the simulation scenarios and the derivation of the underlying truth can be found in Wolbers et al. \citep{wolbers2025shrinkage} and its supplementary materials.

Table \ref{tab:cont-truthtable} summarizes the true treatment effects in the overall population and across all subgroups for the continuous endpoint simulations. The simulation truth was derived based on a synthetic population of 50,000 subjects as described in the main text (Section "Simulation for a continuous endpoint - Data generation").

\begin{table}[!htbp]
\caption{Time-to-event endpoint simulation: True average hazard ratios in subgroups and the overall population for all scenarios.} \label{tab:tte-truthtable}
\centering
\begin{tabular}[t]{lrcccc}
\toprule
Subgroup & Prevalence & Scenario 1 & Scenario 2 & Scenario 3 & Scenario 4\\
\midrule
Overall & 100\% & 0.66 & 0.68 & 0.88 & 0.83\\
$S_{1a}$ & 50\% & 0.66 & 0.68 & 0.74 & 0.58\\
$S_{1b}$ & 50\% & 0.66 & 0.68 & 1.03 & 1.10\\
$S_{2a}$ & 40\% & 0.66 & 0.68 & 0.79 & 0.65\\
$S_{2b}$ & 60\% & 0.66 & 0.68 & 0.95 & 0.95\\
$S_{3a}$ & 20\% & 0.67 & 0.68 & 1.17 & 1.38\\
$S_{3b}$ & 80\% & 0.66 & 0.68 & 0.82 & 0.70\\
$S_{4a}$ & 30\% & 0.66 & 1.00 & 0.88 & 0.82\\
$S_{4b}$ & 30\% & 0.66 & 0.53 & 0.86 & 0.78\\
$S_{4c}$ & 40\% & 0.66 & 0.53 & 0.90 & 0.87\\
$S_{5a}$ & 15\% & 0.67 & 0.68 & 0.96 & 0.97\\
$S_{5b}$ & 15\% & 0.66 & 0.68 & 1.12 & 1.30\\
$S_{5c}$ & 30\% & 0.66 & 0.68 & 0.84 & 0.74\\
$S_{5d}$ & 40\% & 0.66 & 0.68 & 0.80 & 0.68\\
$S_{6a}$ & 40\% & 0.66 & 0.68 & 0.95 & 0.96\\
$S_{6b}$ & 60\% & 0.66 & 0.68 & 0.85 & 0.76\\
$S_{7a}$ & 40\% & 0.66 & 0.68 & 0.91 & 0.89\\
$S_{7b}$ & 60\% & 0.66 & 0.68 & 0.86 & 0.79\\
$S_{8a}$ & 20\% & 0.67 & 0.68 & 0.70 & 0.52\\
$S_{8b}$ & 30\% & 0.66 & 0.68 & 0.99 & 1.01\\
$S_{8c}$ & 50\% & 0.66 & 0.68 & 0.89 & 0.84\\
$S_{9a}$ & 20\% & 0.66 & 0.68 & 0.88 & 0.84\\
$S_{9b}$ & 80\% & 0.66 & 0.68 & 0.88 & 0.82\\
$S_{10a}$ & 20\% & 0.66 & 0.68 & 0.96 & 0.96\\
$S_{10b}$ & 30\% & 0.66 & 0.68 & 0.85 & 0.77\\
$S_{10c}$ & 50\% & 0.66 & 0.68 & 0.87 & 0.80\\
\bottomrule
\end{tabular}
\end{table}

\begin{table} 
\caption{Continuous endpoint simulation: True mean difference in subgroups and the overall population for all scenarios.} \label{tab:cont-truthtable} 
\centering 
\begin{tabular}[t]{lrccc} 
\toprule 
Subgroup & Prevalence & Scenario 1 & Scenario 2 & Scenario 3\\ 
\midrule 
Overall & 100\% & 0.35 & 0.36 & 0.37\\ 
$S_{1N}$ & 62\% & 0.35 & 0.37 & 0.40\\ 
$S_{1Y}$ & 38\% & 0.35 & 0.33 & 0.32\\ 
$S_{2N}$ & 38\% & 0.35 & 0.36 & 0.37\\ 
$S_{2Y}$ & 62\% & 0.35 & 0.36 & 0.37\\ 
$S_{4N}$ & 50\% & 0.35 & 0.36 & 0.37\\ 
$S_{4Y}$ & 50\% & 0.35 & 0.36 & 0.37\\ 
$S_{8N}$ & 71\% & 0.35 & 0.35 & 0.34\\ 
$S_{8Y}$ & 29\% & 0.35 & 0.39 & 0.44\\ 
$S_{11a}$ & 51\% & 0.35 & 0.21 & 0.06\\ 
$S_{11b}$ & 49\% & 0.35 & 0.52 & 0.69\\ 
$S_{14a}$ & 25\% & 0.35 & 0.35 & 0.34\\ 
$S_{14b}$ & 25\% & 0.35 & 0.37 & 0.40\\ 
$S_{14c}$ & 50\% & 0.35 & 0.36 & 0.37\\ 
$S_{17a}$ & 50\% & 0.35 & 0.35 & 0.35\\ 
$S_{17b}$ & 50\% & 0.35 & 0.37 & 0.39\\ 
\bottomrule 
\end{tabular} 
\end{table}

\subsection{Time-to-event endpoint simulation: Results for global models with a normal prior}

\begin{table}[H]
\centering
\begin{threeparttable}
\caption{Time-to-event endpoint simulation (global model with normal prior): Mean (range) of RMSE,  bias, and observed coverage of nominal 95\% confidence or credible intervals across subgroups by simulation scenario.}
\begin{tabular}{llrrrr}
\toprule
Metric & Estimator & Scenario 1 & Scenario 2 & Scenario 3 & Scenario 4\\
\midrule
$\rm{RMSE}\cdot 100$ &    &    &    &    &  \\
  & Global Shrinkage &    &    &    &  \\
  &    HN($\phi=1$) & 14 (13-15) & 16 (13-29) & 16 (13-26) & 19 (13-33)\\
  &    HN($\phi = |\delta_{\mathrm{plan}}|$) & 14 (13-14) & 16 (13-30) & 16 (13-26) & 19 (13-34)\\
  &    HN($\phi = |\delta_{\mathrm{plan}}|/2$) & 14 (13-14) & 16 (13-31) & 16 (13-27) & 19 (13-36)\\
\addlinespace
$|\rm{Bias}|\cdot 100$ &    &    &    &    &  \\
  & Global Shrinkage &    &    &    &  \\
  &    HN($\phi=1$) & 0 (0-1) & 3 (0-24) & 6 (0-20) & 7 (0-26)\\
  &    HN($\phi = |\delta_{\mathrm{plan}}|$) & 0 (0-1) & 3 (1-25) & 6 (0-20) & 7 (0-27)\\
  &    HN($\phi = |\delta_{\mathrm{plan}}|/2$) & 0 (0-1) & 4 (1-27) & 6 (0-22) & 8 (0-31)\\
\addlinespace
Coverage (\%) &    &    &    &    &  \\
  & Global Shrinkage &    &    &    &  \\
  &    HN($\phi=1$) & 97 (95-99) & 95 (74-99) & 94 (83-98) & 93 (82-97)\\
  &    HN($\phi = |\delta_{\mathrm{plan}}|$) & 97 (95-99) & 95 (72-99) & 94 (82-98) & 93 (80-98)\\
  &    HN($\phi = |\delta_{\mathrm{plan}}|/2$) & 97 (94-98) & 94 (65-99) & 94 (80-98) & 92 (75-98)\\
\bottomrule
\end{tabular}
\begin{tablenotes}[flushleft]
\footnotesize
\item Each scenario includes 25 subgroups. Global shrinkage models use a normal prior with a half-normal hyperprior. RMSE and bias were multiplied by 100 to improve readability.
\end{tablenotes}
\end{threeparttable}
\end{table}

\begin{table}[H]
\centering
\begin{threeparttable}
\caption{Time-to-event endpoint simulation (global model with normal prior; scenario 2): Performance of estimators in subgroup 4a where the drug does not work (true $\log(AHR)=0$) and in the subgroup with the worst predicted treatment effect.}\label{tab:tte-table-global-worst}
\begin{tabular}{lccccccc}
\toprule
\multicolumn{1}{c}{ } & \multicolumn{3}{c}{Subgroup 4a} & \multicolumn{4}{c}{Worst Predicted Subgroup} \\
\cmidrule(l{3pt}r{3pt}){2-4} \cmidrule(l{3pt}r{3pt}){5-8}
\makecell[c]{Estimator\\~} & \makecell[c]{RMSE\\~} & \makecell[c]{Bias\\~} & \makecell[c]{Coverage\\(\%)} & \makecell[c]{Accuracy\\(\%)} & \makecell[c]{RMSE\\~} & \makecell[c]{Bias\\~} & \makecell[c]{Coverage\\(\%)}\\
\midrule
\addlinespace
Global Shrinkage &  &  &  &  &  &  & \\
HN($\phi=1$) & 0.29 & -0.24 & 74 & 62 & 0.25 & -0.07 & 85\\
HN($\phi = |\delta_{\mathrm{plan}}|$) & 0.30 & -0.25 & 72 & 62 & 0.25 & -0.08 & 84\\
HN($\phi = |\delta_{\mathrm{plan}}|/2$) & 0.31 & -0.27 & 65 & 62 & 0.26 & -0.11 & 81\\
\bottomrule
\end{tabular}
\begin{tablenotes}[flushleft]
\footnotesize
\item Accuracy is defined as the proportion of simulation runs in which subgroup 4a was correctly identified as having the worst predicted treatment effect. Performance for the worst predicted subgroup is evaluated by comparing the estimate to the true value within the identified subgroup.
\end{tablenotes}
\end{threeparttable}
\end{table}

\subsection{Continuous endpoint simulation: Results for global models with a normal prior}\label{continuous-endpoint-simulation-results-for-global-models-with-a-normal-prior}

\begin{table}[H]
\centering
\begin{threeparttable}
\caption{Continuous endpoint simulation (global model with normal prior): Mean (range) of RMSE,  bias, and observed coverage of nominal 95\% confidence or credible intervals across subgroups by simulation scenario.}\label{tab:cont-table-global}
\begin{tabular}{llrrr}
\toprule
Metric & Estimator & Scenario 1 & Scenario 2 & Scenario 3\\
\midrule
$\rm{RMSE}\cdot 100$ &    &    &    &  \\
  & Global Shrinkage &    &    &  \\
  &    HN($\phi = \sigma_{\mathrm{plan}}$) & 10 (9-10) & 11 (10-14) & 13 (10-19)\\
  &    HN($\phi = |\delta_{\mathrm{plan}}|$) & 10 (9-10) & 11 (10-15) & 13 (10-19)\\
  &    HN($\phi = |\delta_{\mathrm{plan}}|/2$) & 10 (9-10) & 11 (10-15) & 12 (10-20)\\
\addlinespace
$|\rm{Bias}|\cdot 100$ &    &    &    &  \\
  & Global Shrinkage &    &    &  \\
  &    HN($\phi = \sigma_{\mathrm{plan}}$) & 0 (0-0) & 2 (0-9) & 2 (0-13)\\
  &    HN($\phi = |\delta_{\mathrm{plan}}|$) & 0 (0-0) & 2 (0-10) & 2 (0-14)\\
  &    HN($\phi = |\delta_{\mathrm{plan}}|/2$) & 0 (0-0) & 2 (0-10) & 3 (0-16)\\
\addlinespace
Coverage (\%) &    &    &    &  \\
  & Global Shrinkage &    &    &  \\
  &    HN($\phi = \sigma_{\mathrm{plan}}$) & 97 (96-99) & 96 (89-98) & 94 (83-98)\\
  &    HN($\phi = |\delta_{\mathrm{plan}}|$) & 97 (96-99) & 96 (88-98) & 94 (82-97)\\
  &    HN($\phi = |\delta_{\mathrm{plan}}|/2$) & 97 (96-98) & 95 (86-98) & 94 (78-98)\\
\bottomrule
\end{tabular}
\begin{tablenotes}[flushleft]
\footnotesize
\item Each scenario includes 15 subgroups. Global shrinkage models use a normal prior with a half-normal hyperprior. RMSE and bias were multiplied by 100 to improve readability.
\end{tablenotes}
\end{threeparttable}
\end{table}

\begin{table}[H]
\centering
\begin{threeparttable}
\caption{Continuous endpoint simulation (global model with normal prior; scenario 3): Performance of estimators in subgroup 11a which has the worst true efficacy (true treatment effect = 0.06) and in the subgroup with the worst predicted treatment effect.}\label{tab:cont-table-global-worst}
\begin{tabular}{lccccccc}
\toprule
\multicolumn{1}{c}{ } & \multicolumn{3}{c}{Subgroup 11a} & \multicolumn{4}{c}{Worst Predicted Subgroup} \\
\cmidrule(l{3pt}r{3pt}){2-4} \cmidrule(l{3pt}r{3pt}){5-8}
\makecell[c]{Estimator\\~} & \makecell[c]{RMSE\\~} & \makecell[c]{Bias\\~} & \makecell[c]{Coverage\\(\%)} & \makecell[c]{Accuracy\\(\%)} & \makecell[c]{RMSE\\~} & \makecell[c]{Bias\\~} & \makecell[c]{Coverage\\(\%)}\\
\midrule
\addlinespace
Global Shrinkage &  &  &  &  &  &  & \\
HN($\phi= \sigma_{\mathrm{plan}}$) & 0.18 & 0.12 & 85 & 87 & 0.17 & 0.07 & 88\\
HN($\phi = |\delta_{\mathrm{plan}}|$) & 0.18 & 0.12 & 83 & 87 & 0.17 & 0.08 & 86\\
HN($\phi = |\delta_{\mathrm{plan}}|/2$) & 0.19 & 0.14 & 80 & 87 & 0.18 & 0.10 & 84\\
\bottomrule
\end{tabular}
\begin{tablenotes}[flushleft]
\footnotesize
\item Accuracy is defined as the proportion of simulation runs in which subgroup 11a was correctly identified as having the worst predicted treatment effect. Performance for the worst predicted subgroup is evaluated by comparing the estimate to the true value within the identified subgroup.
\end{tablenotes}
\end{threeparttable}
\end{table}

\newpage

\subsection{Continuous endpoint simulation: Results for one-way models with covariate adjustment}

\begin{figure}[!ht]
\centering 
\subfloat[Population estimator and one-way models with and without covariate adjustment.\label{fig:cont-cov-adj-combined-1}]{%
  \includegraphics[width=0.95\linewidth]{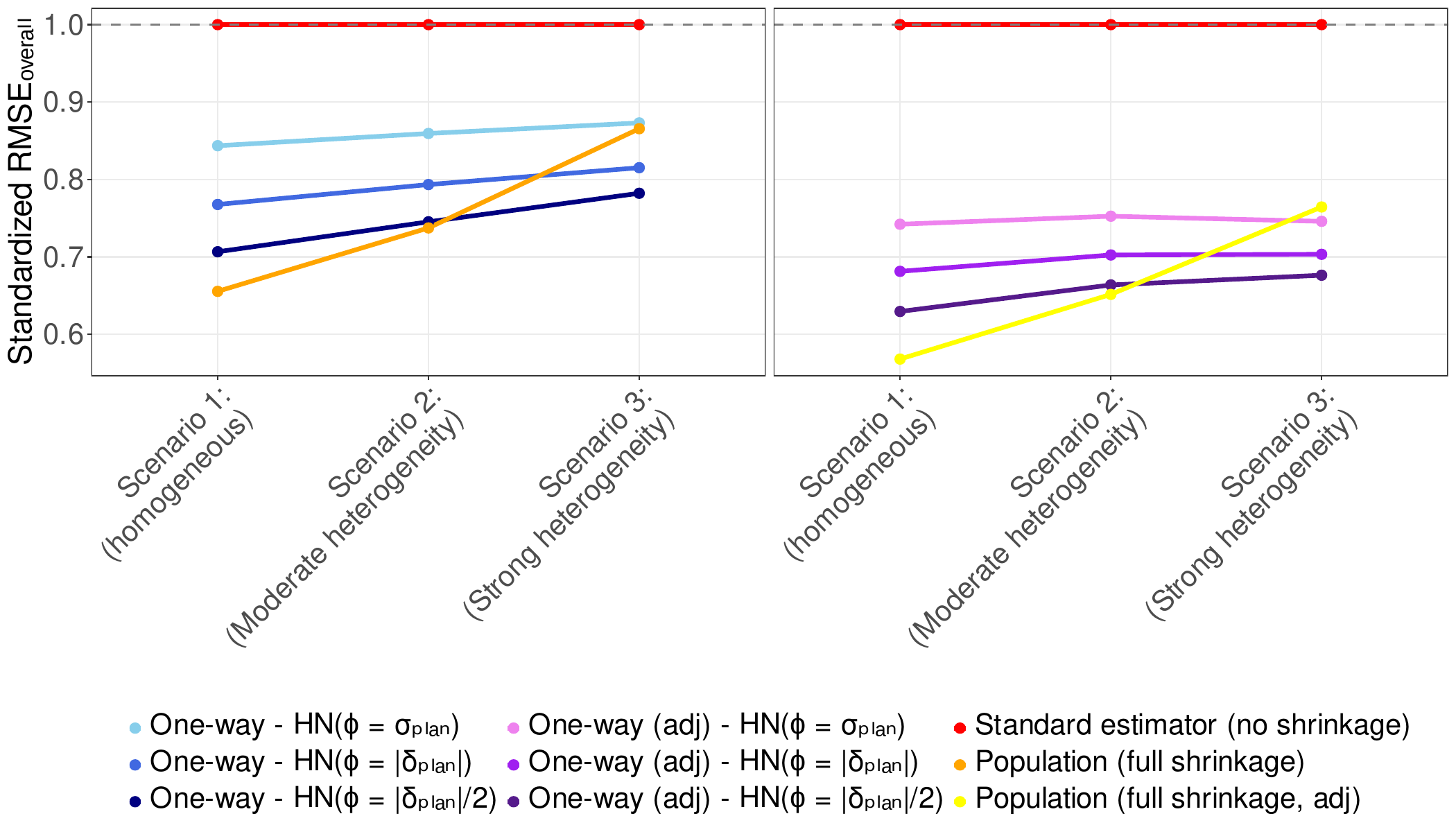}%
}

\subfloat[One-way model with covariate adjustment versus global models with regularized horseshoe prior.\label{fig:cont-cov-adj-combined-2}]{%
  \includegraphics[width=0.95\linewidth]{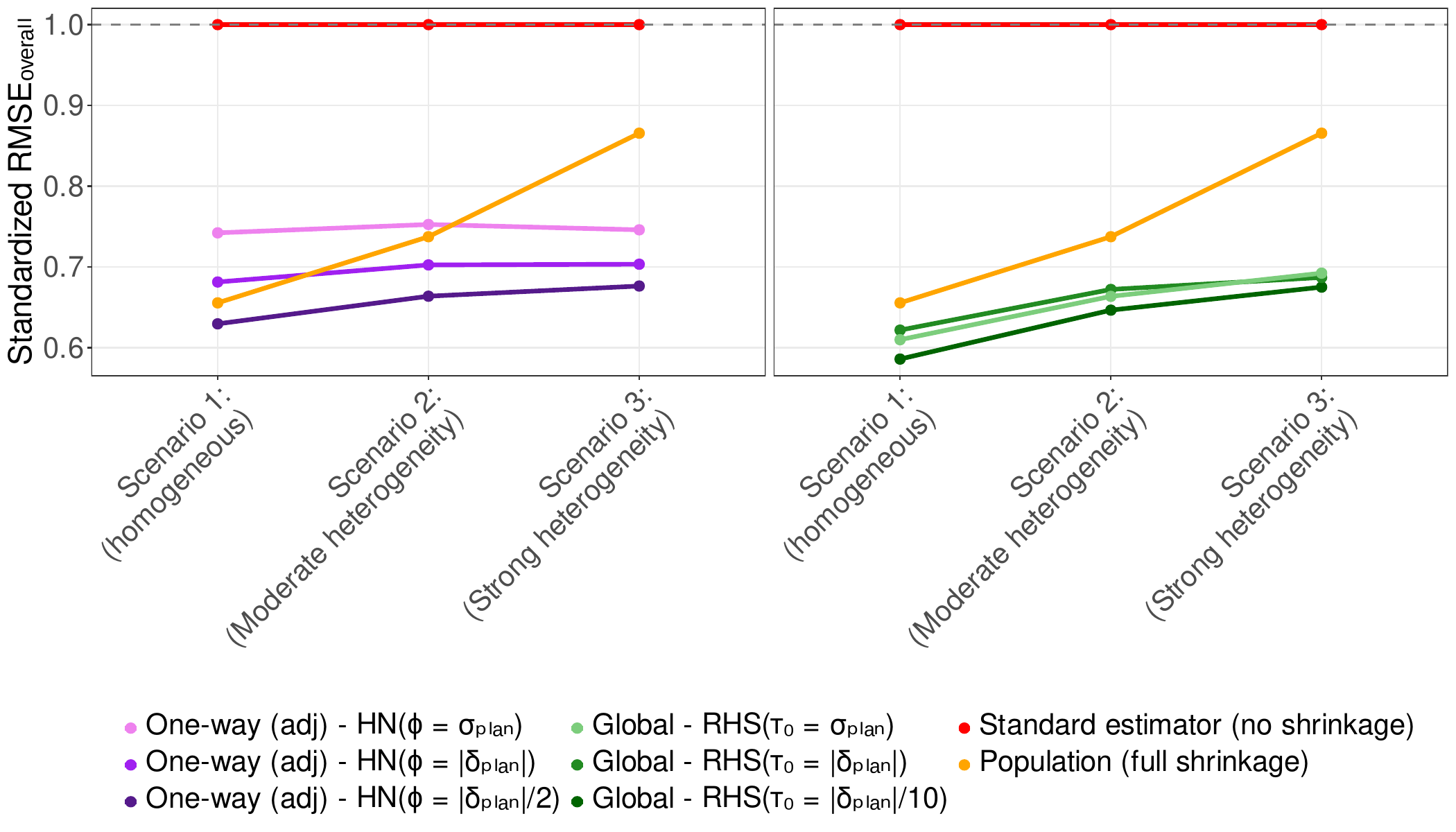}%
}
\caption{Continuous endpoint simulation: RMSE by simulation scenario for covariate-adjusted estimators.}
\label{fig:cont-cov-adj-combined}
\end{figure}

\begin{table}[H]
\centering
\begin{threeparttable}
\caption{Continuous endpoint simulation (covariate-adjusted one-way models): Mean (range) of RMSE,  bias, and observed coverage of nominal 95\% confidence or credible intervals across subgroups by simulation scenario.}
\begin{tabular}[t]{llrrr}
\toprule
Metric & Estimator & Scenario 1 & Scenario 2 & Scenario 3\\
\midrule
$\rm{RMSE}\cdot 100$ &   &   &   &  \\
  & One-way shrinkage &   &   &  \\
  & (covariate-adjusted) & & & \\
  &    HN($\phi = \sigma_{\mathrm{plan}}$) & 12 (10-14) & 12 (11-14) & 13 (11-15)\\
  &    HN($\phi = |\delta_{\mathrm{plan}}|$) & 11 (10-12) & 12 (10-13) & 12 (11-15)\\
  &    HN($\phi = |\delta_{\mathrm{plan}}|/2$) & 10 (9-11) & 11 (10-14) & 12 (10-17)\\
\addlinespace
$|\rm{Bias}|\cdot 100$ &   &   &   &  \\
  & One-way shrinkage &   &   &  \\
  & (covariate-adjusted) & & & \\
  &    HN($\phi = \sigma_{\mathrm{plan}}$) & 0 (0-1) & 1 (0-3) & 1 (0-3)\\
  &    HN($\phi = |\delta_{\mathrm{plan}}|$) & 0 (0-0) & 1 (0-5) & 2 (0-6)\\
  &    HN($\phi = |\delta_{\mathrm{plan}}|/2$) & 0 (0-0) & 2 (0-8) & 2 (0-10)\\
\addlinespace
Coverage (\%) &   &   &   &  \\
  & One-way shrinkage &   &   &  \\
  & (covariate-adjusted) & & & \\
  &    HN($\phi = \sigma_{\mathrm{plan}}$) & 97 (95-98) & 96 (95-98) & 96 (93-97)\\
  &    HN($\phi = |\delta_{\mathrm{plan}}|$) & 97 (96-98) & 96 (94-98) & 96 (92-98)\\
  &    HN($\phi = |\delta_{\mathrm{plan}}|/2$) & 97 (96-99) & 96 (91-98) & 95 (88-98)\\
\bottomrule
\end{tabular}
\begin{tablenotes}[flushleft]
\footnotesize
\item Each scenario includes 25 subgroups. One-way shrinkage models use a normal prior with a half-normal hyperprior. RMSE and bias were multiplied by 100 to improve readability.
\end{tablenotes}
\end{threeparttable}
\end{table}

\begin{table}[H]
\centering
\begin{threeparttable}
\caption{Continuous endpoint simulation (one-way model with covariate adjustment; scenario
3): Performance of estimators in subgroup 11a which has the worst true efficacy (true
treatment effect = 0.06) and in the subgroup with the worst predicted treatment effect.}
\centering
\begin{tabular}{lccccccc}
\toprule
\multicolumn{1}{c}{ } & \multicolumn{3}{c}{Subgroup 11a} & \multicolumn{4}{c}{Worst Predicted Subgroup} \\
\cmidrule(l{3pt}r{3pt}){2-4} \cmidrule(l{3pt}r{3pt}){5-8}
\makecell[c]{Estimator\\~} & \makecell[c]{RMSE\\~} & \makecell[c]{Bias\\~} & \makecell[c]{Coverage\\(\%)} & \makecell[c]{Accuracy\\(\%)} & \makecell[c]{RMSE\\~} & \makecell[c]{Bias\\~} & \makecell[c]{Coverage\\(\%)}\\
\midrule
One-way shrinkage  &  &  &  &  &  &  & \\
(covariate-adjusted) &  &  &  &  &  &  & \\
HN($\phi = \sigma_{\mathrm{plan}}$) & 0.14 & 0.02 & 93 & 87 & 0.15 & -0.03 & 93\\
HN($\phi = |\delta_{\mathrm{plan}}|$) & 0.15 & 0.05 & 92 & 89 & 0.15 & 0.01 & 94\\
HN($\phi = |\delta_{\mathrm{plan}}|/2$) & 0.16 & 0.09 & 89 & 89 & 0.15 & 0.05 & 93\\
\bottomrule
\end{tabular}
\begin{tablenotes}[flushleft]
\footnotesize
\item Accuracy is defined as the proportion of simulation runs in which subgroup 11a was correctly identified as having the worst predicted treatment effect. Performance for the worst predicted subgroup is evaluated by comparing the estimate to the true value within the identified subgroup.
\end{tablenotes}
\end{threeparttable}
\end{table}

\bibliography{SubgroupRefs.bib}

\end{document}